\documentclass[a4paper]{JHEP3}
\usepackage[centertags]{amsmath}
\usepackage{amsfonts,amsmath,amssymb,epsf}
\usepackage{graphicx}

 \def\frac#1#2{{#1\over #2}}

 \def\s{\sqrt}

 \def\de{\partial}

 \def\f {\frac}
 \def\ti{\tilde}
 \def\ap{\alpha}

 \def\no{\nonumber \\}

 \def\frac#1#2{{#1\over #2}}

 \def\s{\sqrt}

 \def\R{\mathbb{R}}

\def\be{\begin{equation}}
\def\ee{\end{equation}}
\def\ba{\begin{eqnarray}}
\def\ea{\end{eqnarray}}

\title{Soliton Stars as Holographic Confined Fermi Liquids}

\author{Jyotirmoy Bhattacharya, Noriaki Ogawa, Tadashi
Takayanagi, Tomonori Ugajin \\ Institute for the Physics and Mathematics of the Universe (IPMU),
 University of Tokyo, Kashiwa, Chiba 277-8582, Japan \\
\email{jyotirmoy.bhattacharya@ipmu.jp},\email{noriaki.ogawa@ipmu.jp},
\email{tadashi.takayanagi@ipmu.jp},\email{tomonori.ugajin@ipmu.jp}}

\abstract{In this paper, we study a holographic dual of a confined
fermi liquid state by putting a charged fluid of fermions in the AdS soliton geometry.
This can be regarded as a confined analogue of electron stars. Depending on the parameters
such as the mass and charge of the bulk fermion field, we found three different phase structures when we change the values of total charge density at zero temperature. In one
of the three cases, our confined solution (called soliton star) is always stable and
this solution approaches to the electron star away from the tip. In both the second and third case, we find a confinement/deconfinement phase transition.
Moreover, in the third one, there is a strong indication that the soliton star decays into an inhomogeneous solution.
We also analyze the probe fermion equations (in the WKB approximation) in the background of this soliton star geometry to confirm the presence of
many fermi-surfaces in the system.
}

\keywords{}
\preprint{IPMU11-0216}

\begin{document}

\section{Introduction}\label{intro}

The AdS/CFT \cite{Maldacena:1997re} has provided us a powerful method
to analyze strongly coupled condensed matter systems. There have been
remarkable progresses on holographic
realizations of Fermi surfaces. For example, the fermions in the
backgrounds of charged AdS black holes
\cite{Liu:2009dm,Faulkner:2009wj,Faulkner:2010zz,Faulkner:2011tm,Iqbal:2011in}
are found to be dual to non-Fermi liquids, while those
in electron star solutions
\cite{Hartnoll:2009ns,Hartnoll:2010gu,Hartnoll:2010ik,Hartnoll:2011fn,Hartnoll:2011dm,Hartnoll:2011pp} (see also \cite{deBoer:2009wk,Arsiwalla:2010bt} for a similar calculation in global AdS)
 are dual to analogues of Landau-Fermi liquids. For closely related examples refer also to e.g.
\cite{Rey:2008zz,Cubrovic:2009ye,Goldstein:2009cv,Gubser:2009qt,Charmousis:2010zz,Goldstein:2010aw}.
In these models, the gravity duals are expected to be dual to deconfined phases of the Fermi surfaces \cite{Sachdev:2010um,Huijse:2011hp,Sachdev:2011wg}.
The properties of fermi surfaces have been usually detected only for their singlet sectors with respect to the $SU(N)$ gauge group.
Recently, however, new identifications of holographic deconfined non-Fermi liquids have been given in \cite{Ogawa:2011bz,Huijse:2011ef,Shaghoulian:2011aa},
where we can confirm the properties of fermi surfaces even for non-singlet sectors by looking at the holographic entanglement
entropy \cite{Ryu:2006bv,Ryu:2006ef,Hubeny:2007xt,Nishioka:2009un} and specific heat. We gave a simple gauge theory model which is expected to have $O(N^2)$ fermi surfaces in the appendix \ref{planarfermion} of this paper.

On the other hand, in realistic condensed matter systems, Fermi surfaces are confined in that there are
no gapless (emergent) gauge bosons. Therefore it is very interesting to construct holographic duals of
confined fermi surfaces as initiated in the works by Sachdev \cite{Sachdev:2011wg,Sachdev:2011ze}. A hard wall model with probe fermions has been
studied in \cite{Sachdev:2011ze}. The purpose of
this paper is to present a fully back-reacted example of confined fermi surfaces by employing the
five dimensional AdS soliton geometry \cite{Witten:1998zw,Horowitz:1998ha}. We assume a
semi-classical description of fermi liquids (Thomas-Fermi approximation) in the AdS soliton
background. We follow the analysis done in the studies of electron stars \cite{Hartnoll:2009ns,Hartnoll:2010gu,Hartnoll:2010ik,Hartnoll:2011dm},
where charged fermionic fluids have been considered in a perturbed Lifshitz geometry \cite{Kachru:2008yh}.
Therefore we will call our solutions {\it soliton stars} in this paper. Also, our model can be regarded as a fermionic analogue of the confined holographic superconductors \cite{Nishioka:2009zj,Horowitz:2010jq}.
Notice that as in the electron star models, we can calculate the (free) energy from our classical gravity dual,
while to obtain the specific heat and entanglement entropy we need to perform one loop calculations in
our gravity dual.\footnote{This is because in our star solutions the (free) energy is the same order as the
curvature of gravity as required by the Einstein equation. To realize this, the chemical potential of the fermi
liquid gets infinitely larger if we take the large $N$ limit. Since the specific heat and (entanglement) entropy
are reduced by the factor of chemical potential compared with the energy, they get much smaller in the large $N$ limit.}

  Below we will find that three different phase structures are possible at zero temperature, depending on the values of two parameters $m$ and $\beta$ of the bulk fermion field.
$m$ denotes the mass of fermion normalized properly. $\beta$ is a parameter related to the fermion charge and number of species of fermions (see appendix \ref{resc}).
Interestingly, in one of the three cases, we find that the soliton star
solutions approach the electron star solutions in major parts of spacetime.
In both the second and third case, we find a first order 
confinement/deconfinement phase transition.
Moreover we found that in the third phase, the soliton star solution gets unstable at a certain
charge density and should decay into a certain new solution which is expected to be inhomogeneous.

In this paper we also analyze the probe fermion equations in the background of this soliton star geometry. In this analysis
we work in the WKB approximation, neglecting the momentum along the compact spatial direction. We use this analysis
to exhibit the presence of a large number of sharp fermi-surfaces in the soliton star system.

This paper is organized as follows. In \S\ref{sec:basic} we present the basic setup of our model.
In \S\ref{sec:persol}, we give perturbative solutions of the soliton star.
In \S\ref{sec:elecstar} we study
electron star solutions in AdS$_5$. In \S\ref{sec:numss}, we numerically analyze the soliton star
solutions. In \S\ref{sec:ratio}, we summarize the behavior of soliton star solutions and study their phase structure.
In \S\ref{sec:nochargess}, we mention the analysis of soliton star without the $U(1)$ charge. In \S\ref{sec:probef}
we consider a probe fermion in the soliton-star background in the WKB approximation and provide evidence of the
existence of fermi surface similar to an ordinary landau fermi liquid. In \S\ref{sec:conclude},
we summarize our conclusions and discuss future problems.

\section{The Basic setup}\label{sec:basic}

We start with the Einstein-Maxwell system defined by the action
\be
S_{EM}=\f{1}{2\kappa^2}\int d^5x \left(R+\f{12}{L^2}-\f{\kappa^2}{4e^2}F^{\mu\nu}F_{\mu\nu}\right). \label{EMth}
\ee
Then we couple this system with fermions. We treat the fermions as a
fermion perfect fluids with the proper equation of state.\footnote{This may be regarded as
 a fermionic analogue of the holographic construction of confined superconductors
studied in \cite{Nishioka:2009zj,Horowitz:2010jq,Chen:2009vz,Pan:2009xa,
He:2010zb,Basu:2010uz,Akhavan:2010bf,Basu:2011yg,Brihaye:2011vk}.}

 In this setup, without losing generality, we can simply set $\kappa=e=L=1$
via the rescaling of parameters of the fermi liquids as explained in the appendix
\ref{resc} and we will do so throughout
in this paper.

\subsection{The Equations of motion}

The Maxwell equation is given by
\begin{equation} \label{mxeq}
 \nabla_{\mu} F_{\nu}^{~\mu} = j^{(f)}_{\nu},
\end{equation}
and the Einstein equations are
\begin{equation} \label{eeq}
 G_{\mu \nu} - 6 g_{\mu \nu} = -\left( F_{\mu \chi} F^{\chi}_{~\nu} - \frac{1}{4} F_{\chi \xi} F^{\xi \chi} g_{\mu \nu}\right) +  T^{(f)}_{\mu \nu},
\end{equation}
where $j^{(f)}_{\nu}$ and $T^{(f)}_{\mu \nu}$ are the charge current and the stress tensor due to the fermions.

Here the coordinates are taken to be $\{t, z, \theta, x, y\}$, where the $\theta$ coordinate is considered to be compact as $\theta\simeq\theta+2\pi$.
The spacetime boundary is $\R^{1,2}\times S^1$ and resides at $z=0$.
The metric and gauge field ansatz for the soliton star are taken to be
\begin{equation}
\begin{split}
 ds^2 &= -f(z) dt^2 + g(z) dz^2 + k(z) d\theta^2
 + \frac{1}{z^2} \left( dx^2 + dy^2\right), \\
 A &= h(z) dt.
\end{split}
\end{equation}

We take the charge current due to the fermions to be of the form
\begin{equation}
 j_{\nu} = \sigma(z) u_{\mu}
\end{equation}
where $\sigma(z)$ is the local charge density of the fermion-fluid and $u_{\mu}$ is the local fluid velocity.
The stress tensor due to the fermions is taken to be that of the ideal fluid form
\begin{equation}
 T^{(f)}_{\mu \nu} = \left( p(z) + \rho(z) \right) u_{\mu} u_{\nu} + p(z) g_{\mu \nu}, \label{EMT}
\end{equation}
where $p(z)$ and $\rho(z)$ are respectively the local pressure and energy density of the fermions.
We choose to work in a frame where the velocity is $u = \{ \frac{1}{\sqrt{f(z)}},0,0,0,0\}$.

It can be shown that the for solving the coupled set of equations \eqref{mxeq} and \eqref{eeq} it is sufficient to solve
the following set of equations.
\begin{align}
p(z) f'(z)+\rho (z) f'(z)-2 \sqrt{f(z)} \sigma (z) h'(z)+2 f(z) p'(z) &= 0 \label{purefer}\\
\frac{z f'(z) \left(z k'(z)-4 k(z)\right)+4 f(z) \left(k(z)-z k'(z)\right)}{4
   z^2 f(z) k(z)}+\frac{h'(z)^2}{2 f(z)}-g(z) p(z)-6 g(z) &= 0 \label{eq1} \\
\frac{f'(z) k'(z)}{2 f(z) k(z)}-\frac{f'(z)}{2 z f(z)}-\frac{3 g'(z)}{2 z
   g(z)}-g(z) p(z)-g(z) \rho (z)-\frac{k'(z)}{2 z k(z)}-\frac{7}{z^2} &=0 \label{eq2}\\
\frac{f'(z)}{2 z^3 f(z) g(z)}-\frac{g'(z)}{2 z^3 g(z)^2}-\frac{k'(z)}{2 z^3
   g(z) k(z)}-\frac{3}{z^4 g(z)} &=   \notag \\
-\frac{f'(z) k'(z)}{4 z^2 f(z) g(z)
   k(z)}+\frac{g'(z) k'(z)}{4 z^2 g(z)^2 k(z)}-\frac{k''(z)}{2 z^2 g(z)
   k(z)}+\frac{k'(z)^2}{4 z^2 g(z) k(z)^2}  & \label{eq3} \\
h'(z)
   \left(\frac{f'(z)}{f(z)}+\frac{g'(z)}{g(z)}-\frac{k'(z)}{k(z)}+\frac{4}{z}
   \right)+2 \sqrt{f(z)} g(z) \sigma (z)-2 h''(z) &=0.        \label{frmmxeq}
\end{align}
Here \eqref{frmmxeq} follows directly from the Maxwell equation \eqref{mxeq}, while the rest are some linear combination of the Einstein equations \eqref{eeq}
and the Maxwell equation. Note that \eqref{purefer} is purely an equation involving the thermodynamic functions characterizing the fermions.

\subsection{The thermodynamic functions of the fermionic fluid} \label{sec:fermionther}

Let the local chemical potential in the tangent frame at a point be given by
\begin{equation}
 \mu (z) = \frac{h(z)}{\sqrt{f(z)}}.
\label{lcp}
\end{equation}
 Assuming the semi-classical approximation of the fermi liquid in the bulk, as has been done for
 electron stars \cite{Hartnoll:2009ns,Hartnoll:2010gu}, the energy density and the charge density to be given by
\begin{align}
 \rho(z) &= \beta \int_m^{\mu(z)} E S(E) dE, \label{rho}\\
 \sigma(z) &= \beta \int_m^{\mu(z)} S(E) dE \label{sig},
\end{align}
where $\beta$ is a constant which is proportional to the number of fermi surfaces.
where $S(E)$ is the density of states. The pressure is given by the Gibbs-Duhem relation
\begin{equation}\label{gd}
 p(z) = - \rho(z) + \mu (z) \sigma(z).
\end{equation}
Note that these expressions for pressure, energy density and charge density in \eqref{rho}, \eqref{sig} and \eqref{gd} automatically satisfy
the equation \eqref{purefer}.

In five dimensions we take the density of states for the fermions to be (we assume that the fermions see the local flat space).
\begin{equation}
 S(E) = E (E^2 - m^2),
\end{equation}
$m$ being the mass of the fermion. With these expression we now have to solve the equations \eqref{eq1}, \eqref{eq2}, \eqref{eq3} and \eqref{frmmxeq} for the
four unknown functions in the metric and the gauge field.

\section{Perturbative solution for the soliton-star}\label{sec:persol}

If we treat the $\beta$ parameter in \eqref{rho} and \eqref{sig} to be small then we can construct our desired soliton star solution perturbatively in $\beta$
about the AdS soliton solution. In this section we present such a construction upto linear order in $\beta$.

We will adopt the following strategy. Inside the star where $\beta$ is non-zero we shall solve the equations linearized in $\beta$ about the AdS soliton solution.
Everywhere inside the star the local chemical potential ($\mu(z)$) will be greater than the mass ($m$) of the fermion. At the radius ($z_r$) of the star the chemical
potential equals the mass of the fermion $\mu (z_r) = m$; in fact  this is the condition that is used to determine the radius of the star. Outside the star $z>z_r$, we
have to solve the vacuum Einstein-Maxwell system with $\beta$ being set to zero. We will then patch up both the solutions at the radius of the star ensuring
continuity of the field strengths.

\subsection{Solution inside the star $z<z_r$}

We can solve for the function $g(z)$ once and for all using the equation \eqref{eq1} after plugging in the pressure as described in \S \ref{sec:fermionther} and obtain
\begin{equation}
 g(z) = -\frac{15 f(z)^{3/2} \left(z \left(k(z) \left(2 z h'(z)^2-4 f'(z)\right)+z
   f'(z) k'(z)\right)+4 f(z) \left(k(z)-z k'(z)\right)\right)}{z^2 k(z)
   \left(-15 \beta  m^4 f(z)^2 h(z)+10 \beta  m^2 f(z) h(z)^3+8 f(z)^{5/2}
   \left(\beta  m^5-45\right)-3 \beta  h(z)^5\right)}.
\end{equation}
Note that this relation is exact in $\beta$. Then we are left with three equations (which follows from \eqref{eq2}, \eqref{eq3}, and \eqref{frmmxeq} respectively) after substitution of the thermodynamic
function according to the prescription of \S \ref{sec:fermionther}. These are three second order equations for the three unknown functions $h(z)$, $k(z)$, and $f(z)$,
then there are 6 initial value (boundary) conditions to be specified.

We solve these equations with the following 6 boundary conditions at the tip of the soliton (which we take to be unity $z_t = 1$):
\begin{enumerate}
 \item There are no corrections to $z_t$ (which is set to 1). This is ensured by the fact that the corrections to the leading order
functional form of $k(z)$ vanishes at $z=1$. Near the tip $k(z) = k_1 (1-z) + {\cal O}(1-z)^2$.
 \item $f(z)$ does not have any logarithmic divergences at the tip $z_t =1$.
 \item $f(1) = 1$.
 \item $h(z)$ does not have any logarithmic divergences at the tip $z_t=1$.
 \item $h(1) = h_0$.
 \item $k_1 = 4 g_{(-1)}$ where $g_{(-1)}$ is given by $g(z) = g_{(-1)}(1-z)^{-1} + {\cal O}(1-z)^0$.
This is required in order that there is no conical deficit of $\theta$ at the tip.
\end{enumerate}

We intend to solve these equation perturbatively in $\beta$ about the AdS soliton geometry.
The the solution inside the star, upto linear order in $\beta$, is given by
\begin{align}
 f_{in}(z) &= \frac{1}{z^2} + \beta ~f^{(1)}_{in}(z) + {\cal O} (\beta^2), \\
 g_{in}(z) &= \left( \frac{1}{z^2 (1 - z^4 )}\right) + \beta ~g^{(1)}_{in}(z) + {\cal O} (\beta^2), \\
 k_{in}(z) &=  \frac{1}{4 }\left( \frac{1}{z^2} - z^2 \right) + \beta ~k^{(1)}_{in}(z) + {\cal O} (\beta^2), \\
 h_{in}(z) &= h_0 + \beta ~h^{(1)}_{in}(z) + {\cal O} (\beta^2).
\end{align}
We must remember that this spacetime exists for $z \leq 1$. The first order functions are complicated and they are explicitly given in appendix \ref{perfunc}.

\subsection{The radius of the star $z_r$}

The radius of the star ($z_r$) is obtained by the following condition
\begin{equation}
 \mu(z_r) \equiv \frac{h_{in}(z_r)}{\sqrt{ f_{in}(z_r)}} = m.
\end{equation}

The radius of the star has the following form
\begin{equation}
 z_r = \frac{m}{h_0} + \beta z_r^{(1)}
\end{equation}
where $z_r^{(1)}$ is the ${ \cal O}(\beta)$ correction to the star radius and is again given explicitly in appendix \ref{perfunc}.

\subsection{The solution outside the star $z<z_r$}

Outside the star we have to solve the same set of equations but now with $\beta$ set to zero.
In this region the $g(z)$ function is expressed in terms of the other functions as follows
\begin{equation}\label{gout}
 g(z) = \frac{z \left(k(z) \left(2 z h'(z)^2-4 f'(z)\right)+z f'(z) k'(z)\right)+4
   f(z) \left(k(z)-z k'(z)\right)}{24 z^2 f(z) k(z)}.
\end{equation}
The zeroth order solution in $\beta$ is identical to the solution inside the star. However the first order solutions are different.
We therefore have
\begin{equation} \label{outpsol}
\begin{split}
 f_{out}(z) &= \frac{1}{z^2} + \beta ~f^{(1)}_{out}(z) + {\cal O} (\beta^2), \\
 g_{out}(z) &= \left( \frac{1}{z^2 (1 - z^4 )}\right) + \beta ~g^{(1)}_{out}(z) + {\cal O} (\beta^2), \\
 k_{out}(z) &= \frac{1}{4 }\left( \frac{1}{z^2} - z^2 \right) + \beta ~k^{(1)}_{out}(z) + {\cal O} (\beta^2), \\
 h_{out}(z) &= h_0 + \beta ~h^{(1)}_{out}(z) + {\cal O} (\beta^2).
\end{split}
\end{equation}
Here $g^{(1)}_{out}(z)$ is expressed in terms of the other functions through the relation \eqref{gout}.
The rest of the functions are given by
\begin{align}
 f^{(1)}_{out}(z) &= \frac{4 F_1-F_2 \log \left(z^4-1\right)}{4 z^2}\\
 k^{(1)}_{out}(z) &= \frac{F_2 \left(z^4-2\right)-F_2 \left(z^4+1\right) \log
   \left(z^4-1\right)+24 \left(K_1 \left(z^4+1\right)+i K_2
   \left(z^4-1\right)\right)}{48 z^2}\\
 h^{(1)}_{out}(z) &= -\frac{1}{4} H_1 \log \left(1-z^2\right)+\frac{1}{4} H_1 \log
   \left(z^2+1\right) + H_2.
\end{align}
Where the integration constants $F_1$,$F_2$, $K_1$, $K_2$, $H_1$ and $H_2$ are determined by condition that
the functions $f(z)$, $g(z)$, and $k(z)$ and their first derivatives are continuous at the star radius $z_r$. This
provides us with 6 conditions to solve for the 6 integration constants. Again their explicit expressions are given
in appendix \ref{perfunc}.

The behavior of the function for a typical values of parameters is shown in fig.\ref{fig:pss}. We use this perturbation
theory as a check on our numerical calculations, which is presented later in \S\ref{sec:numss}. For the values of
values of parameter used in the plot fig.\ref{fig:pss} we find good agreement with the numerical result as shown in
fig.\ref{fig:ssnumc}.

\FIGURE{
\centering
\includegraphics[width=0.7\textwidth]{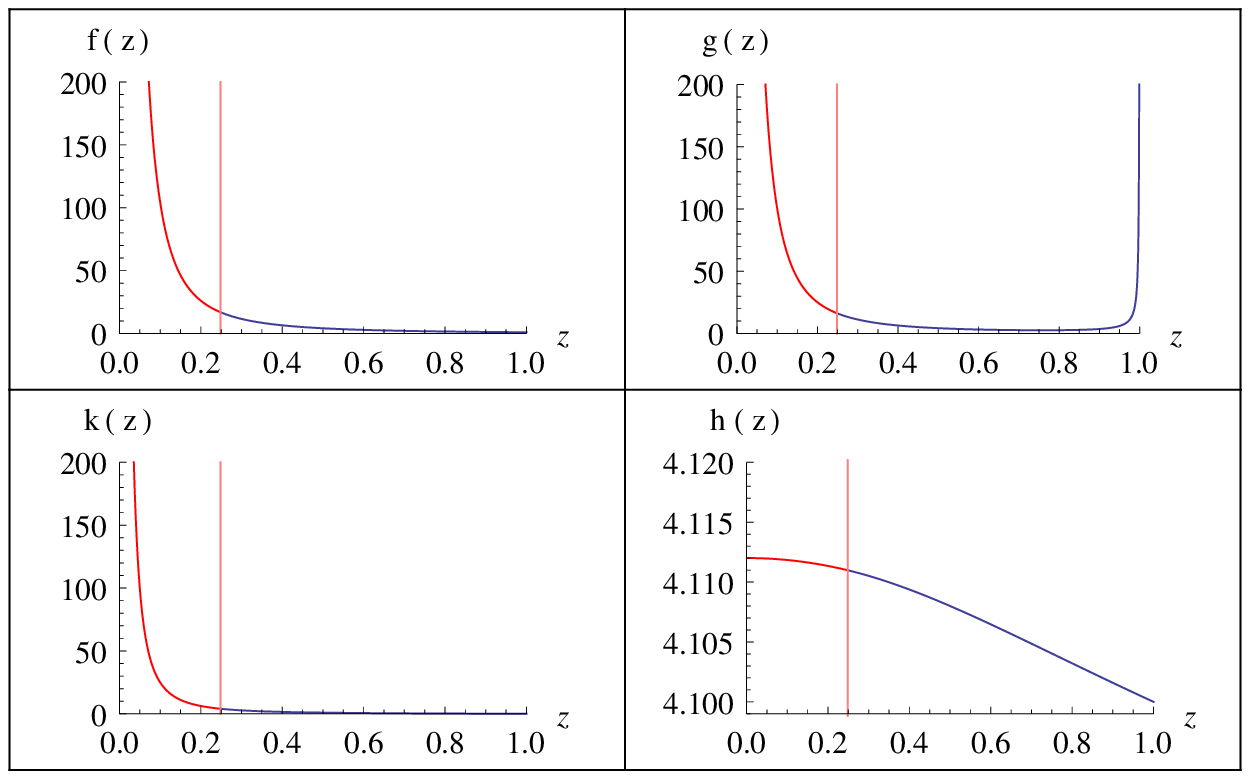}
\caption{Typical perturbative soliton-star ($\beta = 0.001,~m=1,~h_0 = 4.1 $). The vertical red line denotes the radius of the
star which is obtained at $z=0.2484$. The outer solution is denoted by the red line while the inner solution is denoted by the blue line.}
\label{fig:pss}
}

\subsection{Comparison with other solution}

In order to decide which is the dominant solution for a given parameter range, it is necessary to consider all existing solution for that
parameter range. In this paper we primarily focus on the zero temperature solutions and since we fix the temperature (to zero) therefore
we have to work in the canonical ensemble. Also at zero temperature the energy and free energy are same. Hence, we should consider the
energies of the zero temperature solutions in order the decide which is the dominant solution. The one with lower energy is the more stable
solution.

Since our system has conformal symmetry, we have to make this comparison of energy after fixing the scaling symmetry. We fix the scaling symmetry,
by considering only solutions with unit radius of the $\theta$-circle at infinity. We now describe this procedure in details.

We have a metric and gauge field of the form
\begin{equation}
\begin{split}
 ds^2 &= - f(z) dt^2 + g(z)dz^2 + k(z) d\theta^2 + z^{-2} (dx^2 + dy^2), \\
 A &= h(z) dt.
\end{split}
\end{equation}
Our solutions have the following asymptotic form
\begin{equation}\label{asymsolistar}
 \begin{split}
  f(z) &= f_{\infty} z^{-2} - m_{\infty} z^2 + \frac{2}{3} Q z^4+ \dots \\
  g(z) &= \frac{1}{z^2} + \dots \\
  k(z) &= k_{\infty} z^{-2} + k^{(1)}_{\infty} z^2 \dots \\
  h(z) &= \mu - Q z^2 + \dots \\
 \end{split}
\end{equation}

Let us make the following scale transformation on the coordinates
\begin{equation}
 z \rightarrow \frac{1}{c} z; ~~ t \rightarrow \frac{1}{c} t ; ~~ x \rightarrow \frac{1}{c} x ; ~~ y \rightarrow \frac{1}{c} y.
\end{equation}
Under this coordinate change the new parameters of the solution are
\begin{equation}\label{rescaling}
 m'_{\infty} = \frac{m_{\infty}}{c^4};~~ Q' = \frac{Q}{c^3};~~ \mu'=\f{\mu}{c}
 ;~~ f_{\infty}' = f_{\infty};~~ k_{\infty}' = c^2 k_{\infty}.
\end{equation}
If we want to set $k_{\infty}' = 1$ we have to choose
\begin{equation}
 c = \frac{1}{\sqrt{k_{\infty}}}.
\end{equation}

The time-time component of the boundary stress tensor ( $T_{tt}$, which we take to be the energy of our solutions), should scale in the same way as $m_{\infty}$.
Hence the scale invariant quantities that we should compare is $T_{tt} k_{\infty}^2$ at fixed the charge density
\be
 {\cal Q} = \frac{Q}{\sqrt{f_{\infty}}} k_{\infty}^{\frac{3}{2}}.
 \ee
However, practically it is more convenient to look at the dimensionless ratio
 \be
 {\cal R} = \frac{T_{tt}^3}{f_{\infty} Q^4}, \label{ratioe}
 \ee
 at constant ${\cal Q}$.
We therefore compare plots of $ {\cal R}$ versus ${\cal Q}$ for various solutions. The time-time component of the
energy momentum tensor (i.e. the energy density) for the case of out soliton-star is given by
\begin{equation}
 T_{tt} = \frac{3m_{\infty}}{f_{\infty}} + \frac{k^{(1)}_{\infty}}{k_{\infty}}.
\end{equation}
The chemical potential also
provides us a useful information and we will employ the scale invariant
combination
\be
\ti{\mu}=\f{\mu}{\s{f_{\infty}}}\s{k_{\infty}}. \label{norch}
\ee

We can immediately compute this ratio for extremal RN black holes in AdS. The extremal black hole has one parameter less than our soliton star solution and
therefore for it this ratio $ {\cal R}$ is fixed to a definite value independent of ${\cal Q}$. The extremal black hole  is given by the following exact solution
\begin{equation}
 \begin{split}
  f(z) &= z^{-2} - m^{(bh)} z^2 + \frac{2}{3} (Q^{(bh)})^2 z^4,\\
  g(z) &= \frac{1}{z^2 (1 - m^{(bh)} z^4 + \frac{2}{3} (Q^{(bh)})^2 z^6)} , \\
  k(z) &= z^{-2}, \\
  h(z) &= \mu^{(bh)} - Q^{(bh)} z^2. \\
 \end{split}
\end{equation}
with the additional condition $f'(z=z_H) = 0$, $z_H$ being the horizon radius defined by $f(z=z_H) = 0$.

For this solution, using standard AdS/CFT prescription, we get $T_{tt} = 3 m^{(bh)}$ and using this we get
\begin{equation}
 {\cal R} = 81. \label{extbhr}
\end{equation}
for all values of ${\cal Q}$.

\subsection{Validity of the perturbation theory}

In the perturbative solution for the soliton-star $h_0$ may be considered as a parameter of the solution.
Since $h_0$ is the chemical potential at the tip of the soliton, therefore
the value of $h_0$ must be greater than $m$ for the star to exist.
The perturbative solution naively exits for all values of $h_0 \geq m$. However as we go on increasing the
value of $h_0$ the first order correction to various physical quantities increases. Hence, at some value of
$h_0$ the perturbation theory fails to be valid. This point of breakdown of perturbation theory ($h_0^{(p)}$ say)
may be estimated using the following condition
\begin{equation}
 \beta\big|z_r^{(1)}(h_0)\big| \ll \frac{m}{h_0}.
\end{equation}
This condition demands that the first order correction to the radius of the star is always less than the leading order value.

\section{The Electron-star}\label{sec:elecstar}

For comparing the phase structure of various solutions existing in this parameter range we
must also consider so called electron-star solutions. They have fermionic hairs and approach the Lifshitz metric \cite{Kachru:2008yh} in the IR geometry, which have been first constructed in
in 4-dimensions \cite{Hartnoll:2009ns,Hartnoll:2010gu,Hartnoll:2010ik}
(see also recent developments of electron stars in \cite{Puletti:2010de,Hartnoll:2011dm,Cubrovic:2011xm,Edalati:2011yv}). In this section
we construct similar solution in 5-dimensions. These solutions have more symmetry than the soliton star\footnote{For comparison with our soliton-star we will ultimately consider the $\theta$-direction to be
compact which breaks this symmetry. However this is not reflected in the metric which retains
the symmetry between $\theta$,$x$ and $y$ coordinates.}
and the function $k(z)$ is fixed to be
\begin{equation}
 k(z) = \frac{1}{z^2},
\end{equation}
and is no more dynamical.
This choice of the $k(z)$ function automatically solves equation \eqref{eq3} after the thermodynamic functions
according to \S \ref{sec:fermionther} has been substituted. Then we are left with three equations to solve for the three
unknown functions $f(z)$, $g(z)$ and $h(z)$.

\subsection{IR Lifshitz geometry}

Just like the 4-dimensional case, a Lifshitz geometry emerges in the IR of the electron-star which is
independently an exact solution of the system of equations. This solution is given by
\begin{equation}\label{lfgeo}
\begin{split}
 f(z) &= \frac{1}{z^{2 \alpha}} \\
 g(z) &= \frac{g_{L}}{z^2} \\
 h(z) &= \frac{h_{L}}{z^{\alpha}}
\end{split}
\end{equation}
where $g_{L}$ and $h_L$ are given by
\begin{equation}
 \begin{split}
  h_L &=\sqrt{\frac{\alpha -1}{\alpha }}  \\
  g_L &= \frac{1}{60
   \left(\alpha  \left(m^2-1\right)+1\right)^2} \bigg( 16 \sqrt{\alpha -1} \alpha ^{5/2} m^5+5 \alpha ^4
   \left(m^2-1\right)^2+\alpha  \left(80 m^2-53\right) \\ &+\alpha ^3 \left(-5
   m^4-20 m^2+9\right)+\alpha ^2 \left(60 m^4-50 m^2+3\right)+36 \bigg) \\
 \end{split}
\end{equation}
where the Lifshitz coefficient $\alpha$ is implicitly given in terms of the lagrangian parameters
$\beta$ and $m$ by the relation
\begin{equation}\label{betaexp}
\begin{split}
 \beta &= \bigg(720 \sqrt{\alpha -1} \alpha ^{5/2} \bigg) \bigg/ \bigg( 5 \alpha ^4+9 \alpha ^3+3 \alpha
   ^2-53 \alpha  +16 \sqrt{\alpha -1} \alpha ^{5/2} m^5 \\ &+5 \alpha ^4 m^4-5
   \alpha ^3 m^4+60 \alpha ^2 m^4-10 \alpha ^4 m^2-20 \alpha ^3 m^2-50 \alpha
   ^2 m^2+80 \alpha  m^2+36 \bigg)
\end{split}
\end{equation}
A plot of the dynamical exponent $\alpha$ versus $\beta$ for various values of m is shown in fig.\ref{fig:alphabeta}.
Here we take $\alpha$ to be greater than $1$. Also note that if we assume $\alpha$ to be very large then \eqref{betaexp}
takes the following form
\begin{equation}\label{betaexpnd}
 \beta = \frac{144}{\alpha  \left(m^2-1\right)^2}+ { \cal O} \left(\frac{1}{\alpha
   }\right)^{3/2}.
\end{equation}
Remembering that $\beta$ and $m$ are lagrangian parameters which determine $\alpha$, it is clear from
the above expression that $\alpha$ tends to infinity as $\beta$ tends to zero
(from below).
This limit corresponds to the extremal black hole solution with an AdS$_2$ throat.
Also remember that, the equations of motion assume the relation $\mu(z)\ge m$, that is, $h_L\ge m$ from \eqref{lfgeo}.
It gives a bound $0\le m<1$ for the existence of solutions with this form.


\FIGURE{
\centering
\includegraphics[width=0.7\textwidth, angle=-90]{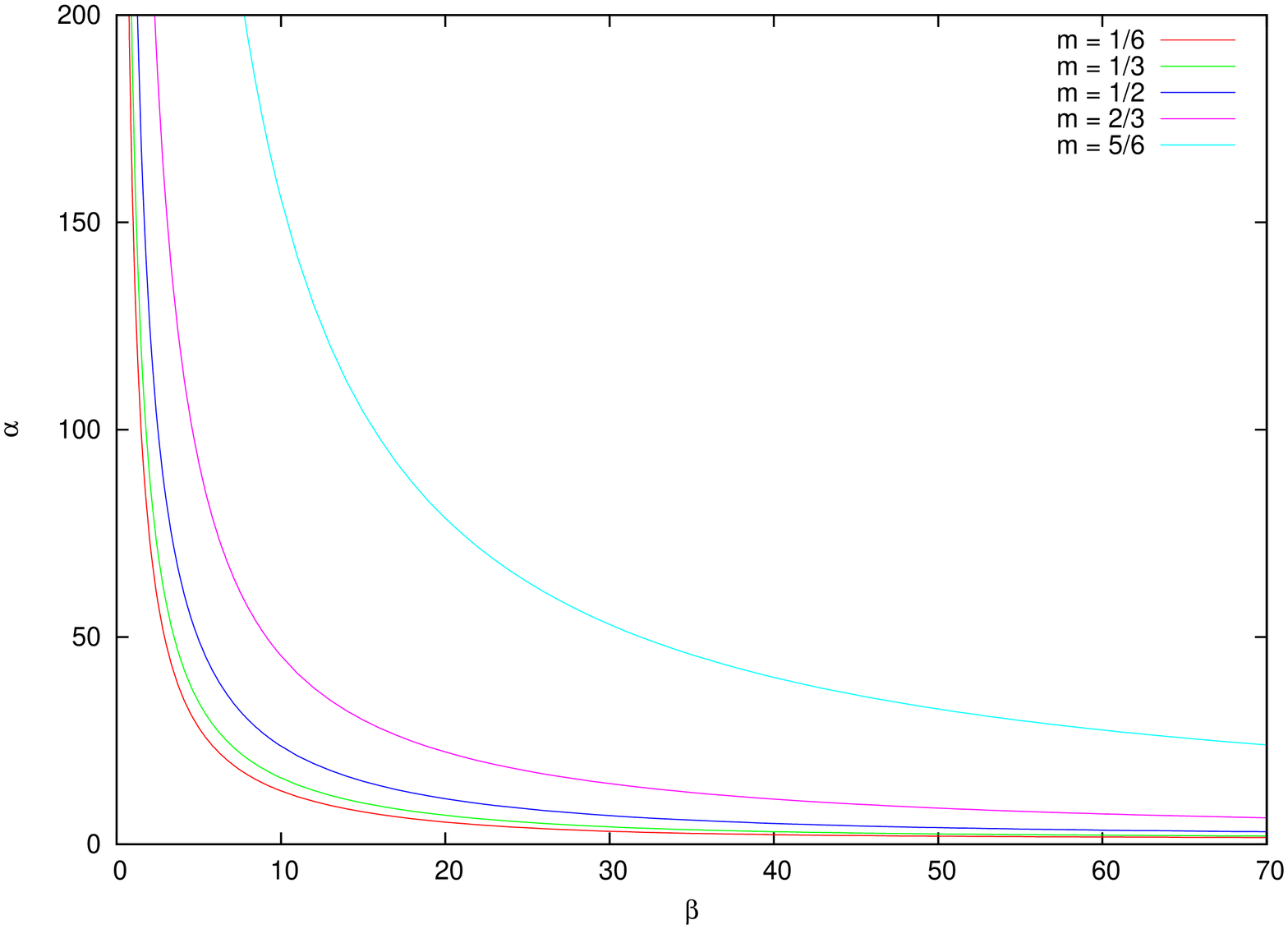}
\caption{A plot of the dynamical exponent $\alpha$ versus $\beta$ for various values of m.}
\label{fig:alphabeta}
}

As emphasized previously this is an exact solution of the Einstein solution.
However to obtain our electron star solution we shall perturb away from this solution keeping it in the deep IR.
We shall then use this perturbation result to set the boundary condition for our numerical solution.

\subsection{Perturbation away from the IR Lifshitz geometry}

We now consider a perturbation away from the IR Lifshitz geometry in \eqref{lfgeo} in the following way
\begin{equation}\label{lfgeoper}
\begin{split}
 f(z) &= \frac{1}{z^{2 \alpha}} \left(1 + \sum_{i} f_{es}^{(i)} z^{i \gamma} \right) \\
 g(z) &= \frac{g_{L}}{z^2} \left( 1 + \sum_{i} g_{es}^{(i)} z^{i \gamma} \right) \\
 h(z) &= \frac{h_{L}}{z^{\alpha}} \left( 1 + \sum_{(i)} h_{es}^{(i)} z^{i \gamma} \right).
\end{split}
\end{equation}
Here $f_{es}^{(1)}$ is a parameter of the solution, whose magnitude is unphysical due to a
scaling symmetry. However, it is very important for it to have a negative sign for the solution
to exist. Using the scaling symmetry we set its value to -1.

Also clearly for the perturbation theory to be meaningful we have to consider the negative root for
$\gamma$. The first correction to the Lifshitz solution is given by,
{\small
\begin{equation}
\begin{split}
 g_{es}^{(1)} &= \frac{\gamma  \left(\gamma -2 \alpha ^2 \left(m^2-3\right)+\alpha
   \left(-\gamma +(\gamma +1) m^2+1\right)-7\right)}{\gamma ^2+\gamma +\alpha
   ^3 \left(7-3 m^2\right)+\alpha ^2 \left(-\gamma +(\gamma +3)
   m^2+10\right)+\alpha  \left(-\gamma ^2+\left(\gamma ^2+\gamma -12\right)
   m^2+19\right)-36}\\
 h_{es}^{(1)} &= \bigg( -3 \gamma ^2+9 \gamma +\alpha ^4 \left(7-3 m^2\right)+\alpha ^3 \left(6
   m^2+3\right)+3 \alpha ^2 \left(-\gamma +(\gamma -5) m^2+3\right) \\ & +\alpha
   \left(3 \gamma ^2-6 \gamma -3 \left(\gamma ^2-3 \gamma -4\right)
   m^2-55\right)+36\bigg) \bigg/ \bigg(2 (\alpha -1) \left(-\gamma ^2-\gamma +\alpha ^3 \left(3
   m^2-7\right) \right. \\& \left. +\alpha ^2 \left(\gamma -(\gamma +3) m^2-10\right)+\alpha
   \left(\gamma ^2-\left(\gamma ^2+\gamma -12\right) m^2-19\right)+36\right) \bigg), \\
\end{split}
\end{equation}
}
where $\gamma$ has three roots
\begin{equation}
\begin{split}
 \gamma &= 3+ \alpha, \\&\frac{1}{2 \left(\alpha -\alpha  m^2-1\right)} \bigg(\alpha ^2+2 \alpha -\alpha ^2 m^2-3 \alpha  m^2 -3
\pm \bigg( 9 \alpha ^4-32
   \alpha ^3+94 \alpha ^2-128 \alpha +9 \alpha ^4 m^4 \\ &-30 \alpha ^3 m^4+25
   \alpha ^2 m^4-18 \alpha ^4 m^2+62 \alpha ^3 m^2-126 \alpha ^2 m^2+82
   \alpha  m^2+57 \bigg)^{\frac{1}{2}}
\end{split}
\end{equation}
Here $3+ \alpha$ is the trivial root. As pointed out earlier we have to choose the manifestly negative value of $\gamma$ (for reasonable
values of the parameters) and this is given by
\begin{equation}
\begin{split}
 \gamma =& \frac{1}{2 \left(\alpha -\alpha  m^2-1\right)} \bigg(\alpha ^2+2 \alpha -\alpha ^2 m^2-3 \alpha  m^2 -3
- \bigg( 9 \alpha ^4-32
   \alpha ^3+94 \alpha ^2-128 \alpha \\ &+9 \alpha ^4 m^4 -30 \alpha ^3 m^4+25
   \alpha ^2 m^4-18 \alpha ^4 m^2+62 \alpha ^3 m^2-126 \alpha ^2 m^2+82
   \alpha  m^2+57 \bigg)^{\frac{1}{2}}.
\end{split}
\end{equation}

Even in this case we were able to go to higher order but avoid reporting them here to avoid clumsy expressions.

\subsection{Numerical solution for the Electron star}

Now we use the analytical perturbative solution about the IR Lifshitz geometry in \eqref{lfgeoper} to set the initial conditions for
our numerical solution at point in $z$ in the deep IR. After numerically carrying the solution forward we find the radius of
the electron star when the bulk local chemical potential becomes equal to mass of the fermion. As a check of our calculations
we have verified that the value of the radius of the star is independent of the value of the IR point chosen to put the
boundary condition. Also we have to remember that there exists an upper bound of the fermion mass which follows from \eqref{betaexpnd}
and is given by $0 \leq m \leq 1$.

The solution outside the star radius is again a vacuum solution (with more symmetry than in case of the soliton star)
and it is possible to write this solution down analytically
\begin{equation}\label{esoutsol}
 \begin{split}
  f(z) &= \chi^2 z^{-2} - m^{(es)} z^2 + \frac{2}{3} (Q^{(es)})^2 z^4,\\
  g(z) &= \frac{\chi^2}{z^2 (1 - m^{(bh)} z^4 + \frac{2}{3} (Q^{(es)})^2 z^6)} , \\
  h(z) &= \mu^{(es)} - Q^{(es)} z^2. \\
 \end{split}
\end{equation}
Note that this solution is essentially the black hole solution with an additional
parameter $\chi$ which is usually set to unity by a scaling symmetry in the usual
black hole solution. However in the present case, we have already used this symmetry
in the inner solution to set $f_{es}^{(1)} = -1$. The four constants in
\eqref{esoutsol}, $\chi$,  $m^{(es)}$, $\mu^{(es)}$ and $Q^{(es)}$ are determined
by matching with inner solution at the radius of star.

\FIGURE{
\centering
\includegraphics[width=0.7\textwidth]{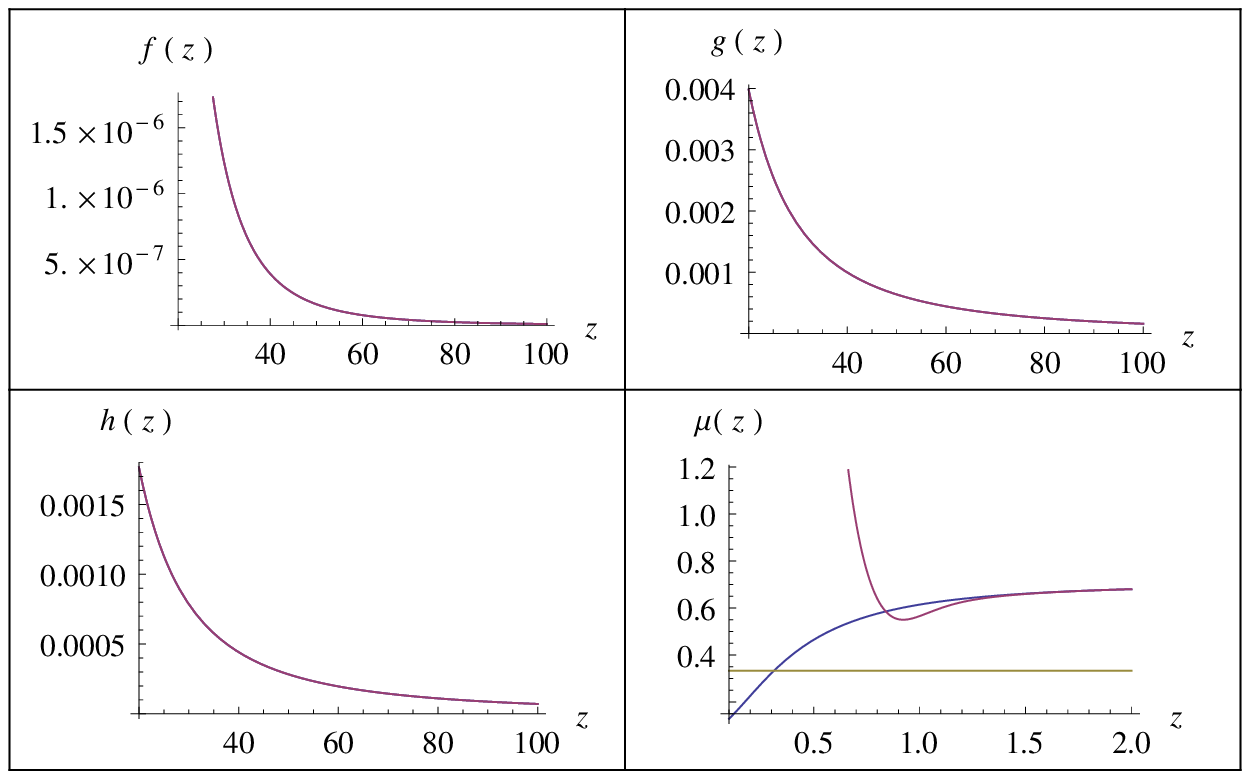}
\caption{Typical solution inside the electron-star for $m=1/3, \beta = 70.43$, same values of parameter
for the numerical soliton star plot. Note that the blue line is the actual numerical solution while the
red line is the perturbative solution about the IR Lifshitz geometry. The numerical solution
matches very well with the perturbative solution in the IR region but begins to break down when z becomes
${\cal O}(1)$. This is visible in the plot of the chemical potential $\mu(z)$. Here we see that
the perturbative solution is unable to capture the radius of the electron star and therefore to
do so we have to use the numerical solution. The yellow line in the $\mu(z)$-plot denotes the value of the
mass $m=1/3$.}
\label{fig:es}
}

In Fig.\ref{fig:es} we plot the numerical functions inside the star radius
for a particular value of the parameter. This electron star solution
just like the black hole solution does not have the extra parameter
present in the case of the soliton star. Therefore, like the black hole,
in this case, the energy density to charge density ratio ${\cal R}$ is
determined to be a number independent of
${\cal Q}$. A plot of ${\cal R}$  versus $\beta$ for various choice  of $m$ is shown in fig.\ref{fig:ratiobeta}

\FIGURE{
\centering
\includegraphics[width=0.7\textwidth, angle=-90]{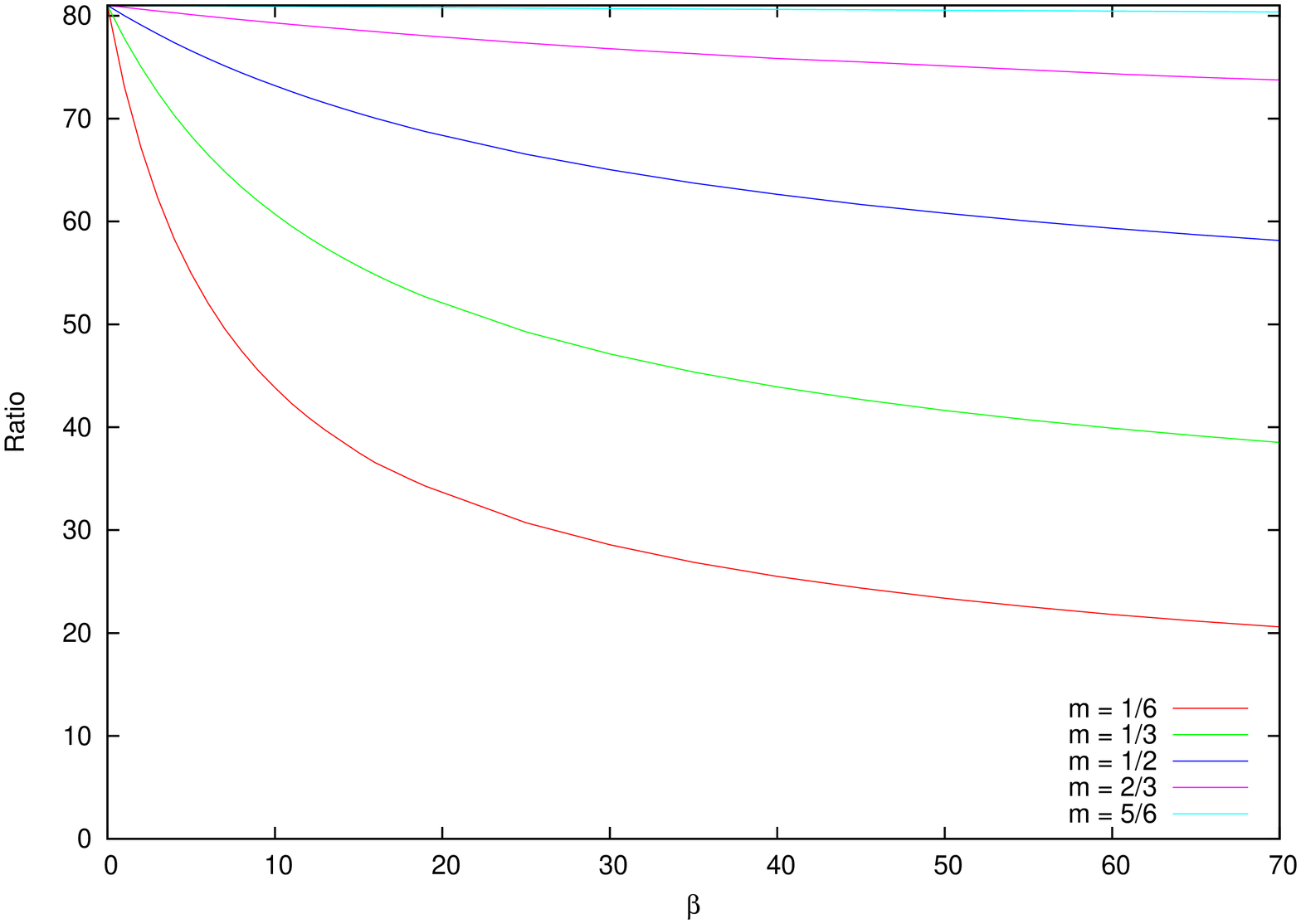}
\caption{A plot of the energy to charge ratio ${\cal R}$ versus $\beta$ for various choice  of $m$ for the electron star.}
\label{fig:ratiobeta}
}

\section{Numerical solution for the soliton-star} \label{sec:numss}
Now we go back to the soliton star solutions. To understand its behavior for generic
values of the parameters, we need to resort to a numerical approach.
In order to set the initial conditions at the tip of the star for our numerical calculation we first find an analytical solution very near the tip of the soliton.

\subsection{ Solution near the tip of the soliton}\label{sec:tipsol}

Let us choose our coordinate scales such that the tip of the soliton occurs at $z=1$ just as in the case of the perturbative solution in \S \ref{sec:persol}.
Also similar to the perturbative solution we use the invariance under scaling of the time coordinate to set the function $f(z)$ to 1 at the tip $z=1$.
We then make the following ansatz for our dynamical functions near the tip $z=1$:
\begin{equation}\label{tipsol}
 \begin{split}
  f(z) &= 1 + f_1 (1-z) 
+ {\cal O} (1-z)^2, \\
  g(z) &= \frac{g_{(-1)}}{1-z} + g_0  
+ {\cal O} (1-z), \\
  k(z) &= 4 g_{(-1)} (1-z) + k_2 (1-z)^2 
+ {\cal O} (1-z)^2, \\
  h(z) &= h_0 + h_1 (1-z) 
+  {\cal O} (1-z)^2. \\
 \end{split}
\end{equation}
Here $h_0$ continues to be a parameter of the solution and has the same interpretation of being the
value of the local bulk chemical potential at the tip of the soliton. Now substituting this ansatz
back into the equations and solving them order by order of $(1-z)$,
we can determine the unknown constants in \eqref{tipsol} in terms of $h_0$
(and the lagrangian parameters which in this case is $\beta$ and $m$). The result of this procedure is given in
appendix \ref{tipfunc}. We have actually gone one higher order in $(1-z)$ in order to put the initial conditions
for the numerical calculation to greater precision. However since the constants appearing
in that order are very large and complicated therefore we refrain from reporting them here.

\subsection{Numerical solution}

We use the perturbative solution in \S \ref{sec:tipsol} to set the initial conditions for the soliton star near the tip
of the soliton. We then numerically evolve the solution forward. Just like in the perturbative solution in \S \ref{sec:persol},
we determine the radius of the star ($z_r$) from the following condition
\begin{equation}
 \mu(z_r) = \frac{h(z_r)}{\sqrt{f(z_r)}} = m.
\label{sszr}
\end{equation}
Then outside the radius of the star we have the vacuum equations (with $\beta =0$).
Unlike the electron star case given by \eqref{esoutsol}, we do not know the analytic solution even outside the star. Therefore we again solve these equations numerically
with the boundary conditions being provided by the inner solution at the radius of the star. This solution is then
continued to the asymptotic boundary from where the energy density $m_{\infty}$, charge density $Q$ and the $\theta$-circle radius are read off using a
fitting function of the form \eqref{asymsolistar}.

A plot of the dynamical functions, for example, for a particular value $\beta=70.43$ and $m=1/3$ of the parameters is presented in Fig:\ref{fig:ssnum}.
This value of $\beta$ is chosen so that the Lifshitz
solution has the dynamical exponent $\ap=2$. As a check on our numerics we compare our numerical solution with the perturbative solution
plotted in fig.\ref{fig:pss} in \S\ref{sec:persol} for small value of $\beta$ ($\beta = 0.001$). The comparison plots are shown in fig.\ref{fig:ssnumc} which shows that there is
considerable agreement between the results.

\FIGURE{
\centering
\includegraphics[width=0.7\textwidth]{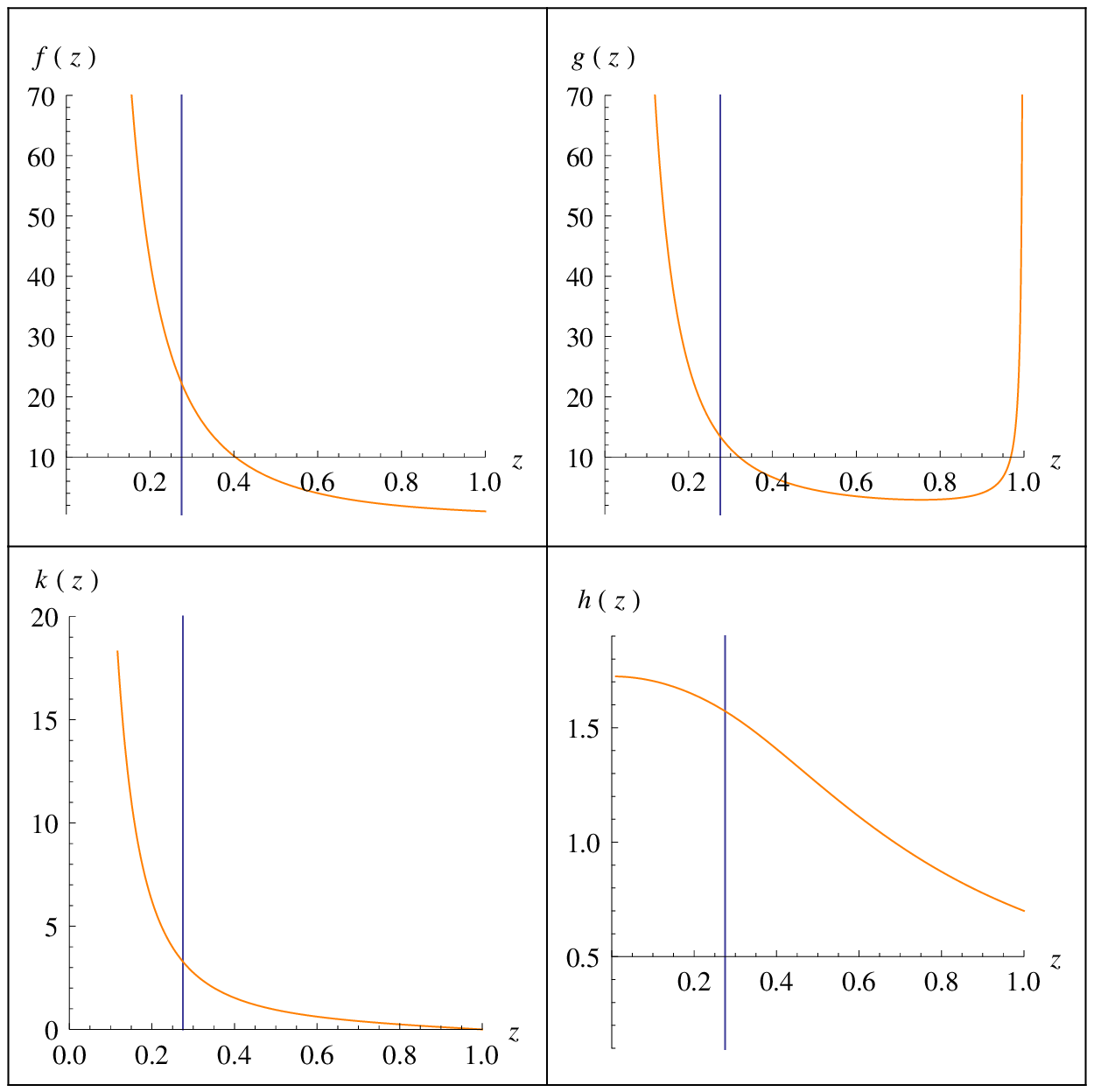}
\caption{Typical numerical soliton-star ($m=1/3,~\beta= 70.43, ~h_0 = 0.7$). The blue line represents the radius of the star at $z=0.275$.}
\label{fig:ssnum}
}

\FIGURE{
\centering
\includegraphics[width=0.7\textwidth]{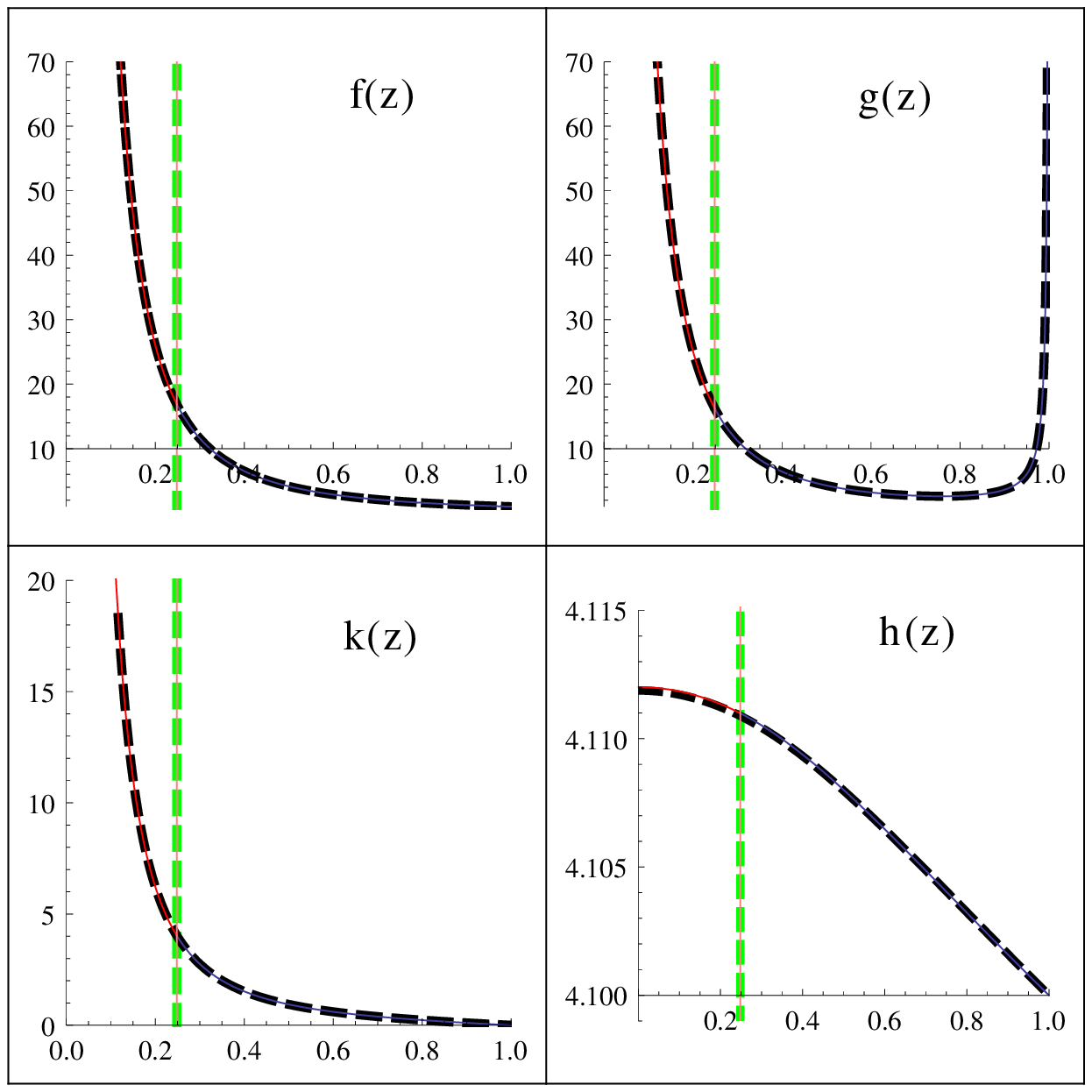}
\caption{In this plot we compare the numerical solution obtained in this section with the perturbative solution in $\beta$. The value of
parameters that we use are $\beta = 0.001,~m=1,~h_0 = 4.1 $. The dotted lines (green and black) represent the numerical solution while
the continuous lines represent the perturbative solution. Numerically
the value of the radius is obtained to be $z=0.2483$ which compares well with the perturbative value of the radius $z=0.2484$. }
\label{fig:ssnumc}
}

\subsection{Bounds on $h_0$} \label{h0bound}

As pointed out earlier the parameter $h_0$ must be greater than the mass $m$ of the constituent fermions.
This is because $h_0$ is simply the local chemical potential at the tip of the soliton and for any
fermion condensation to take place it has to be greater than the mass of the constituent fermions.
Therefore $h_0$ clearly has a lower bound for the solution to exist.

More importantly, it turns out that $h_0$ also has an upper bound. There are two reasons for the presence of upper bound. One of them can be understood from the
fact that the denominator of $g_{(-1)}$ in \eqref{tipsol} blows up at this bound. In fact we loose
control over the numerical calculation as we approach this point even before we reach it. From
\eqref{tipsolconst1} this denominator is given by
\begin{equation}
 {\cal D} (h_0) = 9 \beta  h_0^5-10 \beta  h_0^3 m^2-15 \beta  h_0
   m^4+16 \left(\beta  m^5-45\right).
\end{equation}
The upper bound is given by the real root of the polynomial ${\cal D} (h_0)$. This upper bound for $h_0$ is denoted by $h^*_0$ below. At $h_0=h^*_0$ the metric gets singular (or divergent) at the tip $z=1$. For example it is given by $h^*_0=1.053$ for $\beta=70.43$ and
$m=1/3$.

Another reason of the upper bound is because when $\beta$ is larger than some value $\beta_c$ (this is $\beta_c\simeq 20$ for $m=1/3$), the solution develops a singularity somewhere between the tip and the AdS boundary%
\footnote{
It would be no problem, if there were to exist a point satisfying
\eqref{sszr}, before the singularity.
However it is not the case here.
}
 before $h_0$ reaches $h^*_0$. This upper bound of $h_0$ is denoted by $h^c_0(<h^*_0)$. For example it is given by $h^c_0=0.774$ for $\beta=70.43$ and
$m=1/3$. Thus for $\beta>\beta_c$, the true upper bound is given by $h^c_0$, while for $\beta\leq\beta_c$ it is given by $h^*_0$. Note that this phenomena only happens for
the mass range $0\leq m< 1$. In other words, for $m\geq 1$, we always have $\beta_c=\infty$.

The behavior of the soliton star solution changes when $\beta$ crosses $\beta_c$.
If $\beta>\beta_c$, as $h_0$ approaches $h^c_0$, the charge density ${\cal Q}$ gets infinitely large
and the star radius $z_r$ gets closer to the AdS boundary $z=0$.
This means that we can realize a very large star. On the other hand, if $\beta<\beta_c$, when $h_0$ approaches $h^*_0$, the charge density
remains finite and approaches to a certain value ${\mathcal Q}_{max}$. In this case, the star radius also stays a certain range and does not get closer to the AdS boundary. The behavior of the star radius $z_r$ in these limits as a function of $\beta$ is plotted at $m=1/3$ in Fig.\ref{fig:yval}. It will be an interesting future problem to analyze this transition in more detail.

\FIGURE{
\centering
\includegraphics[width=0.7\textwidth]{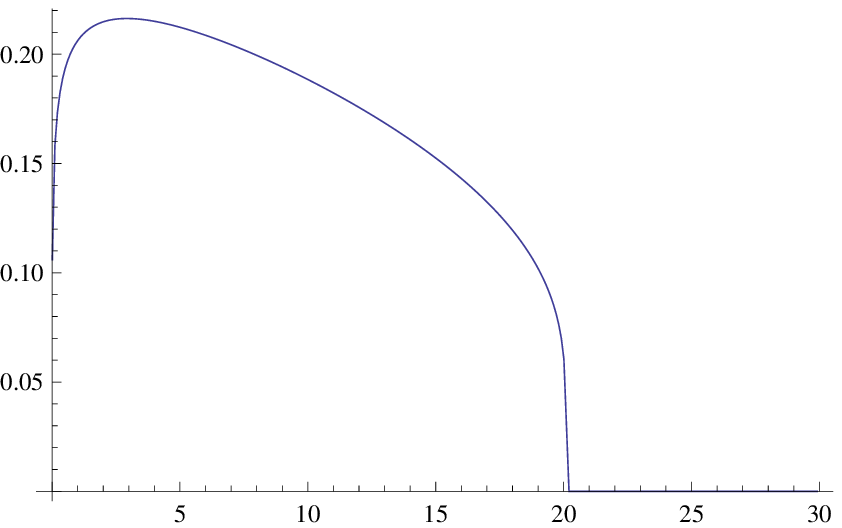}
\caption{The star radius $z_r$ as a function of $\beta$ at $m=1/3$.}
\label{fig:yval}
}
%
\subsection{Thermodynamical Stability}\label{tins}

It is also helpful to examine the relation between the
chemical potential  $\ti{\mu}$ and the charge density ${\mathcal Q}$. The thermodynamical
stability requires
\be
\f{\de\ti{\mu}}{\de {\cal Q}}=\f{\de^2 E}{\de {\cal Q}^2}\geq 0,  \label{ineqs}
\ee
where $E$ is the total energy of our system. If this derivative gets negative,
then inhomogeneous configurations where the charge is concentrated on some particular regions
can be energetically favored over the homogeneous one. As we will show explicit numerical
results in section \ref{compa} later, the inequality \eqref{ineqs} is satisfied at least
for a small charge density. Generically, when the charge density is larger than a specific value ${\mathcal Q}_c$, the condition \eqref{ineqs} is violated. For such a value of ${\mathcal Q}$ the soliton star solution
becomes thermodynamically unstable and moreover, a perturbative instability is also well expected in the spirit of the Gubser-Mitra conjecture \cite{Gubser:2000ec,Gubser:2000mm}

In the case  $0\leq m< 1$, for $\beta>\beta_c$, no violation of \eqref{ineqs} is observed,
while $\beta<\beta_c$, there exists a point ${\mathcal Q}={\mathcal Q}_c$ where it starts to be violated.
On the other hand, in the case $m\geq 1$, for any $\beta$ we find the instability point
${\mathcal Q}_c$. We will explain how these instabilities affect the phase structure of our
system later in section \ref{compa}.

\section{Comparison of Energy for various solutions}\label{sec:ratio}

\FIGURE{
\centering
\includegraphics[width=0.8\textwidth]{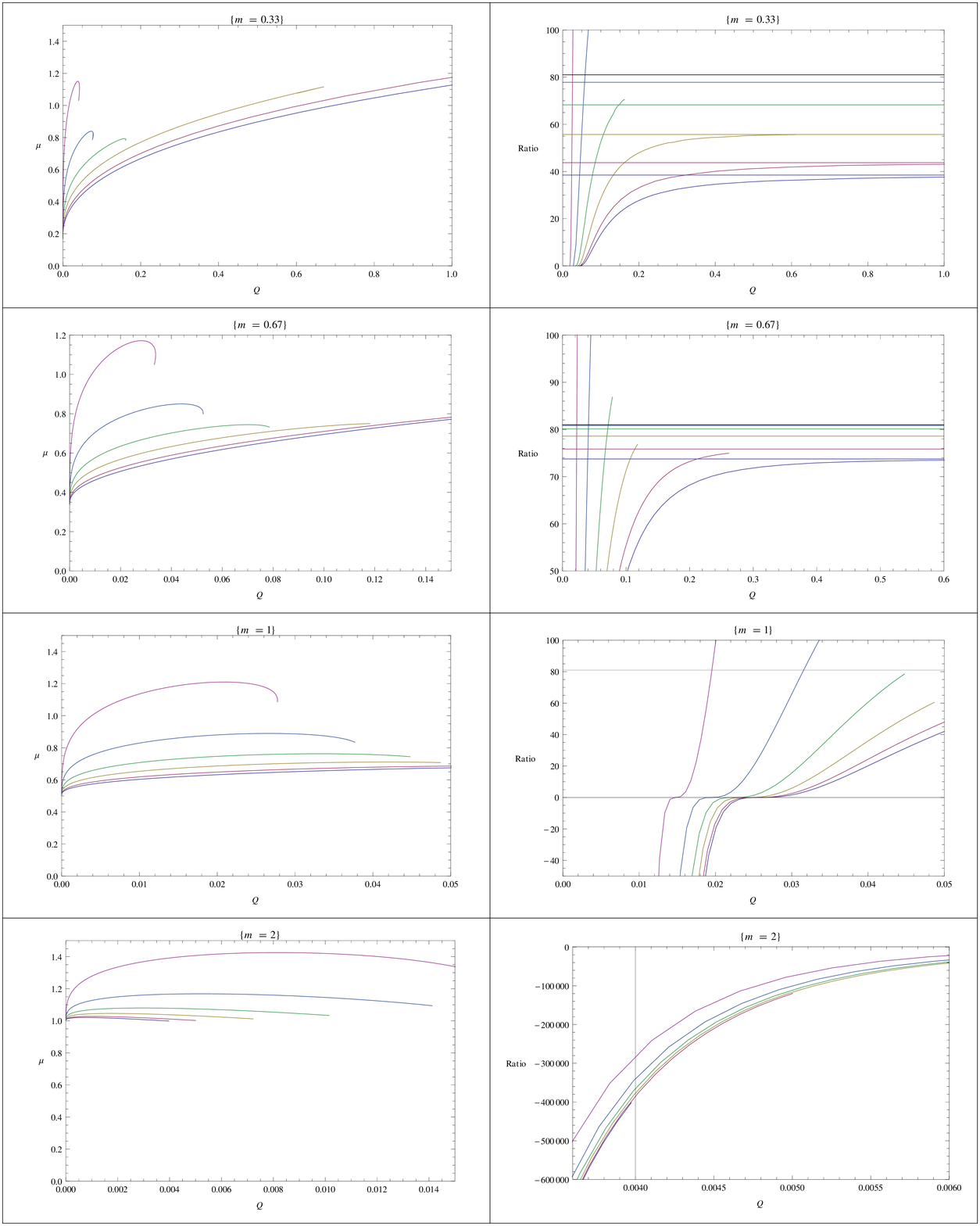}
\caption{The plots of the chemical potential $\ti{\mu}$ and ratio ${\mathcal R}$ as functions of the charge density $Q$. The value of $m$ is indicated at the top of each graph. Each graph consists of six colored curves which correspond to (from the up to the down) $\beta=0.1$ (red), $\beta=1$ (blue), $\beta=5$ (green),
$\beta=15$ (yellow), $\beta=40.7$ (red), $\beta=70.4$ (blue). We also plotted the values of
${\mathcal R}$ for the electron stars with these values of $\beta$ as well as the that for the
extremal black hole.}
\label{fig:compare}
}

\FIGURE{
\centering
\includegraphics[height=10cm]{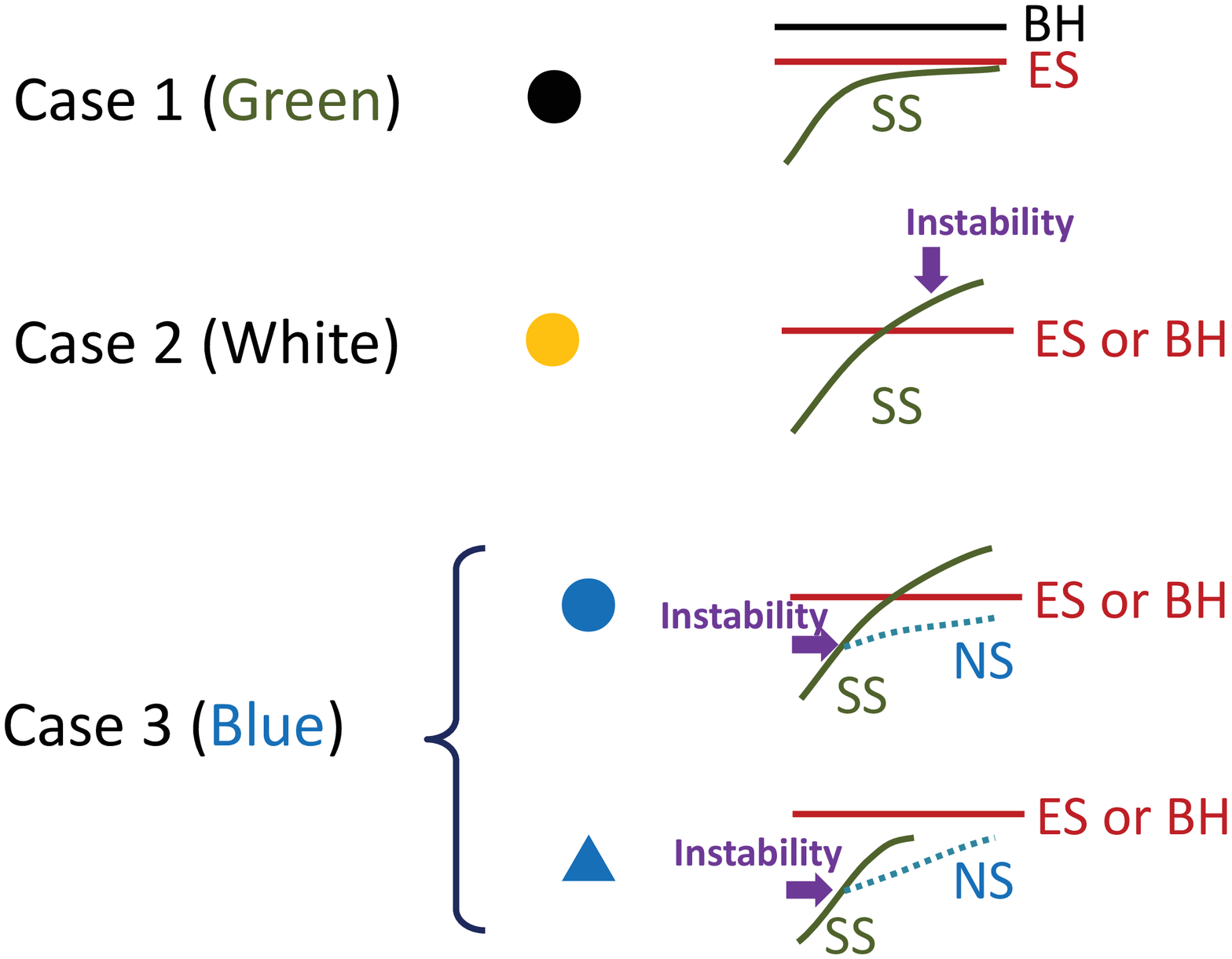}
\caption{The three possible phase structures for various values of $\beta$ and $m$.
In the right figures we show the schematic graph of the ratio ${\mathcal R}$ as a function of the charge density ${\mathcal Q}$ for
the extremal charged black hole (BH), the electron star (ES) and the soliton star (SS).
We also indicate the point where the thermodynamical instability defined by $\f{\de \mu}{\de Q}<0$ starts.}
\label{fig:threephase}
}

\FIGURE{
\centering
\includegraphics[height=8cm]{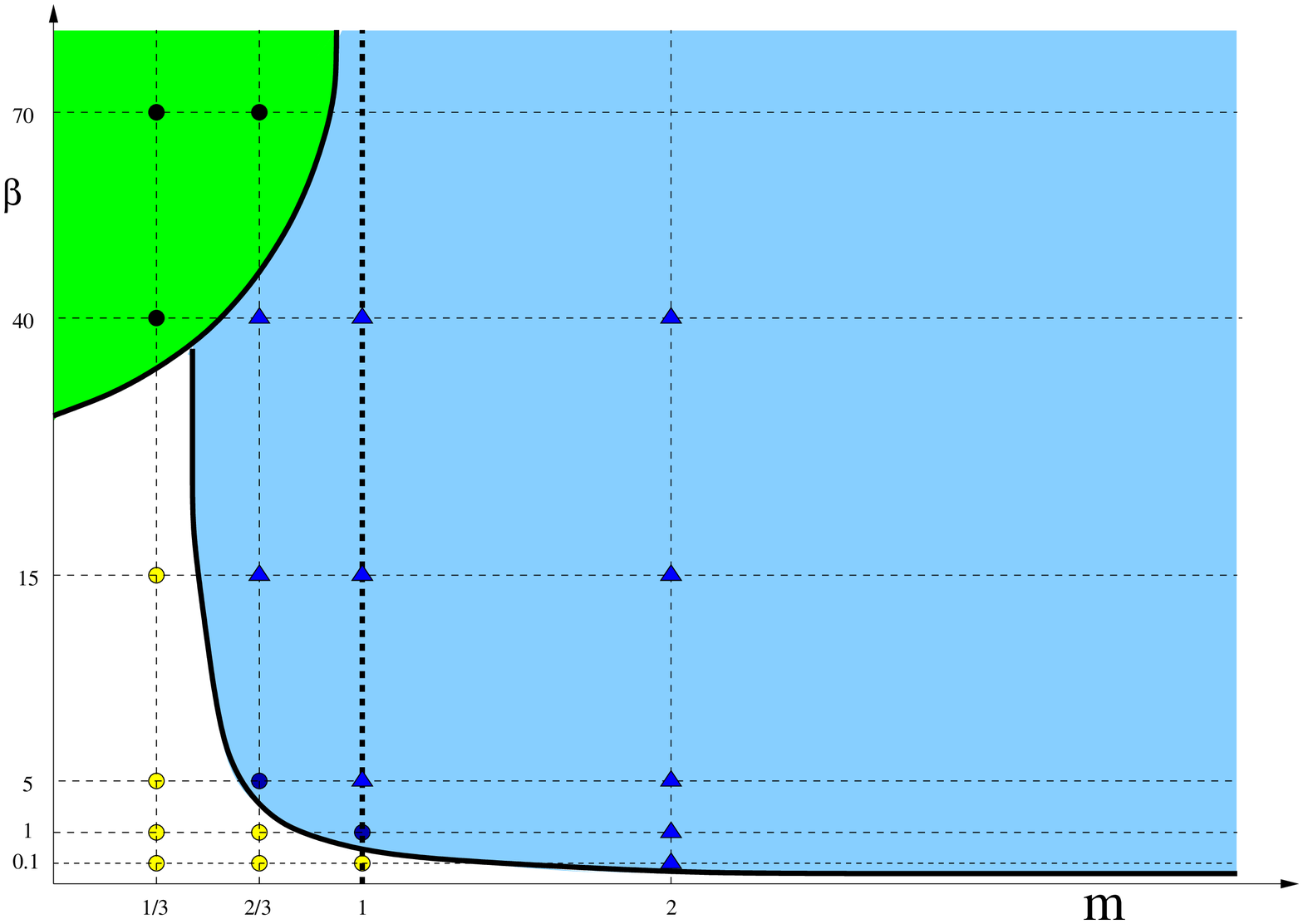}
\caption{The Schematic phase diagram for various $\beta$ and $m$. The green region, the white region and the blue region correspond to the Case 1, Case 2 and Case 3. Each point with a small
colored circle and triangle, which is defined by the previous picture, represents an example we performed an explicit numerical calculation.}
\label{fig:phased}
}

As follows from our previous arguments, for a given generic value of $(\beta,m)$, we will have
three different solutions at zero temperature:
(i) a soliton star (SS), (ii) an electron star (ES)and (iii) an extremal
charged black hole (BH), which are all smooth solutions in our Einstein-Maxwell-Fermion system.
Notice that the electron star only exists when $0\leq m<1$, while the others exist for any
$\beta$ and $m$. The physically relevant solution at zero temperature is the one with the
lowest energy or equally the one with the lowest value of the ratio ${\mathcal R}$ \eqref{ratioe}
for a fixed charge density ${\mathcal Q}$.
Notice that extremal charged black holes have always the same value ${\mathcal R}=81$ \eqref{extbhr}. An electron star takes a fixed value ${\mathcal R}$ which does not depend on ${\mathcal Q}$ for a given
$(\beta,m)$. On the other hand, a soliton star has a non-trivial function
${\mathcal R}={\mathcal R}({\mathcal Q})$, which also depends on $(\beta,m)$. As we discussed
in section \ref{tins}, the behavior of the chemical potential $\ti{\mu}$, normalized
as in \eqref{norch}, is very important to know the thermodynamical stability. We plotted the
${\mathcal R}$ and $\ti{\mu}$ as a function of ${\mathcal Q}$ for the several values of $(\beta,m)$.

\subsection{Phase Structure}\label{compa}

Based on the behaviors of  ${\mathcal R}$ and $\ti{\mu}$, we can classify the phase structure at
zero temperature into three cases (simply called case 1, case 2 and case 3) as explained in Fig.\ref{fig:threephase}. We present the schematic diagram in Fig.\ref{fig:phased} to show
which case corresponds to a given value of $(\beta,m)$. Before we explain each case, we would like to note that the soliton star is the stable solution with the lowest energy if the charge density is enough small. This is obvious from the fact that the AdS soliton is the lowest energy solution to
the vacuum Einstein equation \cite{Horowitz:1998ha}.

The case 1 is characterized by the condition $\beta>\beta_c$, where $\beta_c$ is
the lower bound of $\beta$ which allowed the limit of infinitely large charge density ${\mathcal Q}\to \infty$ as discussed in section \ref{h0bound}. In this case, as we can see from Fig.\ref{fig:compare},
the soliton star solution always has the lowest energy (or equally ${\mathcal R}$). The condition
\eqref{ineqs} is always satisfied and there is no thermodynamical instability.
Moreover, we notice an intriguing fact that the value ${\mathcal R}$ of the soliton star approaches that of the electron star in the large charge density limit. Actually, as we show in the section \ref{slimit}, the solution star solution itself approaches to that of the electron star in this limit. This is also consistent with the fact that the case 1 occurs only in the mass range $0\leq m<1$.

On the other hand, if $\beta<\beta_c$, the system corresponds either case 2 or case 3. In these
cases, we always encounter the upper bound of charge density ${\mathcal Q}_c$ above which \eqref{ineqs} is violated and the system gets thermodynamically unstable. The difference between the case 2 and 3 comes from the detail of how the instability starts. In the case 2, the instability occurs
at a enough large value of charge density and the energy density of solition star gets larger than that of either the electron star (for $0\leq m<1$) or the extremal black hole (for $m\geq 1$) before the onset of this instability. Therefore, in the case 2, there is a first order phase transition from the soliton star into either the electron star or the extremal black hole.
This can be regarded as a deconfinement/confinement transition.

In the case 3, the instability takes place when the energy of soliton star is still
the lowest one. In this case, the solition star should decay into a new solution (NS) in order to
have a sensible phase structure as is clear from Fig.\ref{fig:threephase}. It is possible
that for a much larger value of charge density the electron star or extremal black hole is finally favored. This appearance of the new blanch will be related to a perturbative instability of
solition star solution for a large enough charge density. As suggested from the thermodynamical instability, it is natural to expect that the new solution is inhomogeneous with respect to the $x,y$ directions. It will be a very important future problem to work out this conjectured perturbative instability and new solution.

In summary, the ground state for each of the three cases changes in the following way
as we increase the charge density ${\mathcal Q}$:
\ba
&&\mbox{Case 1 : Soliton star}  \no
&&\mbox{Case 2: Soliton star$\to$ Electron star ($0\leq m <1$) or Extremal BH ($m\geq1$)}\no
&&\mbox{Case 3: Soliton star$\to$ New solution ($\to$ Electron Star ($0\leq m <1$) or Extremal BH ($m\geq1$))}.\nonumber
\ea
The phase transition found in the case 2 can be regarded as a confinement/deconfinement phase
transition. In the case 1, this transition goes smoothly and eventually the solition star
degenerates with the electron star in the large charge density limit as we mentioned. In the case 3,
we first encounter an instability against inhomogeneous perturbations.

\subsection{Electron star as a limit of soliton star}  \label{slimit}

In this section we explain that the soliton star approaches electron star solution
when both $\beta$ and the charge density are large, as we already mentioned in section \ref{compa}. The fact is intuitively clear since the bulk chemical potential  $\mu(z)$ monotonically
decreases from the tip to the AdS boundary and bounded by $h_{0}$.
If we pack more and more fermions,  $\mu(z)$ tends to be constant and in
that case corresponding solution becomes the Lifshitz one which coincides with the IR region of electron star.
This also explains the reason why the
ratio ${\mathcal R}$ of solition star approaches that of the electron star for large $\beta$ and
large $Q$ as observed in the previous section.

In  figure \ref {fig:stoe} we compare the part of the metric $g(z)$ of soliton star solution
with nearly maximum value of $h_{0}$ and that of electron star when $\beta=70.43$ and $m=\frac{1}{3}$. To compare
both solutions we use rescaling (\ref {rescaling}) so that radii of both stars  are same. Although  in the region very close to tip, both solutions look different
but if we move toward the boundary they eventually coincide.

\FIGURE{
\centering
\includegraphics[width=0.7\textwidth]{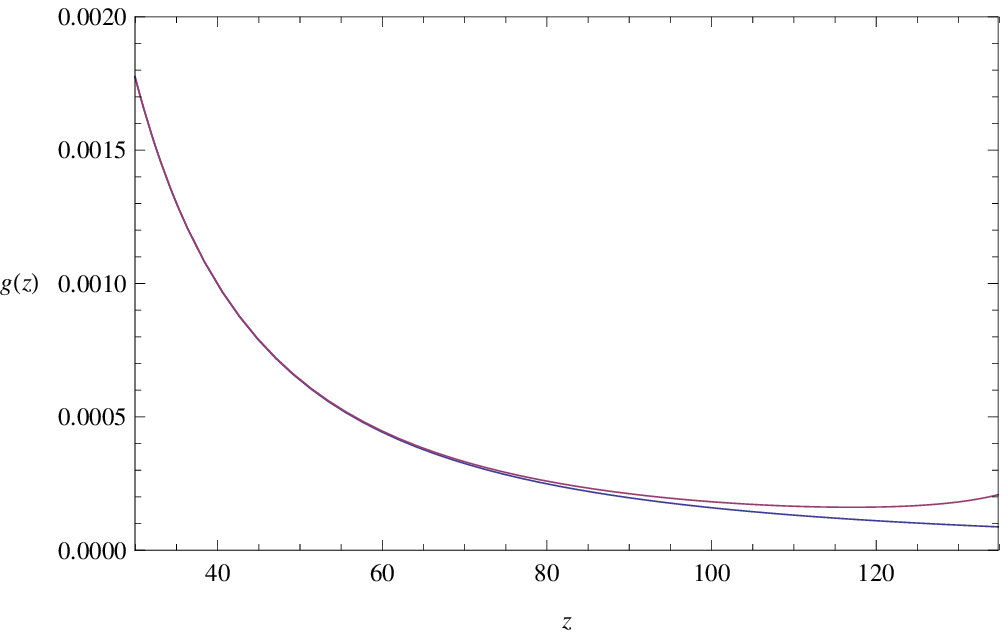}
\caption{Comparison of the solutions in $m=1/3, \beta=70.43,h_{0}=0.7744$ case. The red line is $g[z]$ of soliton star and the blue line is $g[z]$ of electron  star. The tip of
soliton star is located at  $z=149.8$}
\label{fig:stoe}
}

\section{Soliton Star without Charge}\label{sec:nochargess}

It is worth mentioning that Einstein fermionic fluid system without any
$U(1)$ gauge field also admit soliton star solutions. In this section we consider these solutions. 
One can regard the solutions as the poincare patch versions of the AdS degenerate star solutions considered in \cite{Arsiwalla:2010bt}. In this system, equation of motions are given by just setting $h(z)$ to zero in Einstein equations in (\ref{eq1})(\ref{eq2})(\ref{eq3}), and $\mu(z)$ appearing at (\ref{rho}),(\ref{sig}) is given by
\begin{equation}
\mu(z)=\frac{\epsilon_{F}}{\sqrt{f(z)}},
\label{cp_woc}
\end{equation}
where $\epsilon_{F}$ is arbitrary constant (and regarded as  Fermi energy at the tip) so that the stress tensor of the fluid $T^{\mathit{fluid}}_{\mu \nu}$ satisfy $\bigtriangledown^{\mu}T^{\mathit{fluid}}_{\mu \nu}=0$.
Alternatively, a $U(1)$ gauge transformation results in $h(z)\equiv \epsilon_F$,
and in that frame \eqref{cp_woc} coincides with \eqref{lcp}.

One can numerically solve the equations as in previous case. In figure \ref{fig:ssg} we plot the energy density of the star $T_{tt}/f_\infty $ as function of $\epsilon_{F}$ in the case of $m=1,\beta=1$.

\FIGURE{
\centering
\includegraphics[width=0.7\textwidth]{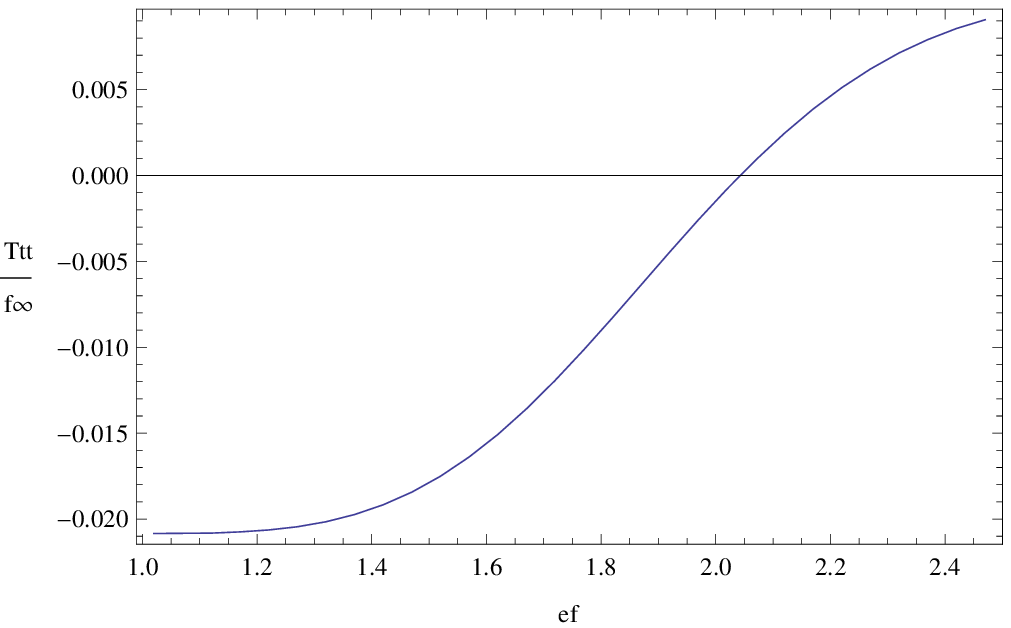}
\caption{Plot of $T_{tt}/f_\infty$ as function of $\epsilon_{F}$ in the case of $m=1,\beta=1$ }
\label{fig:ssg}
}

\section{Probe fermions in the soliton star geometry}\label{sec:probef}

In this section we shall consider a probe fermion in the background soliton star geometry obtained in \S\ref{sec:numss}.
In this study we shall closely follow the analysis of \cite{Faulkner:2009wj,Hartnoll:2011dm}. We wish to show (in a WKB
approximation) that there are many well defined fermi-surface momentum for our soliton star. Because of the
confining nature of the background geometry the gapless excitations in the system are only those associated with these fermi-surfaces.
Thus our system is a very good holographic realization of the standard fermi-liquid. Further using a
recent result in \cite{Iqbal:2011bf} we argue that Luttinger theorem trivially holds for our system.

\subsection{Presence of Fermi-surfaces from a probe fermion analysis}

The background soliton star geometry consists of a metric of the form
\begin{equation}
 ds^2 = -f(z) dt^2 + g(z) dz^2 + k(z) d\theta^2 + \frac{1}{z^2} \big( dx^2 + dy^2 \big),
\end{equation}
together with a gauge field
\begin{equation}
 A = h(z) dt.
\end{equation}
Since we want to deal with fermions therefore we make the following choice of the tetrad for the metric under consideration
\begin{equation}\label{tetrad}
 e^{\bar{t}}_t =  \sqrt{f(z)}; ~~e^{\bar{z}}_z =  \sqrt{g(z)};~~e^{\bar{\theta}}_\theta =  \sqrt{k(z)};~~e^{\bar{x}}_x = \frac{1}{z};~~e^{\bar{y}}_y =  \frac{1}{z},
\end{equation}
where the bar signifies tangent space indices.

In this background geometry we consider a spinor field with the following action
\begin{equation}\label{diracac}
 S = \int d^5x \sqrt{-g} ~i\bigg( \bar{\psi} \Gamma^M {\mathcal D}_{M} \psi -m~ \bar{\psi} \psi \bigg) + \dots
\end{equation}
where the dots in the above action signify appropriate counterterms so as to make the boundary theory well defined.
Here the covariant derivative is given by
\begin{equation}
 {\mathcal D}_M = \partial_M + \frac{1}{4} \omega_{abM}\Gamma^{ab} - i q A_M .
\end{equation}
Note that here we have used small letters (like $a,b$) to denote the tangent space indices, where as capital letters (like $M$) to denote
bulk coordinate indices. Here, $m$ and $q$ are respectively the mass and charge of the probe fermion.
Also, we have used
\begin{equation}
 \Gamma^{ab} = [\Gamma^a, \Gamma^b],
\end{equation}
and $\omega_{abM}$ here is the spin connection. For the choice of tetrad in \eqref{tetrad} the non-zero components of the spin connection is
\begin{equation}
 \omega_{{\bar{t}}{\bar{z}}} = \frac{f'(z)}{2 \sqrt{f(z) g(z)}} dt; ~~\omega_{{\bar{\theta}}{\bar{z}}} = \frac{k'(z)}{2 \sqrt{k(z) g(z)}} d\theta;
~~ \omega_{{\bar{x}}{\bar{z}}} = -\frac{1}{z^2 \sqrt{g(z)}} dx;~~ \omega_{{\bar{y}}{\bar{z}}} = -\frac{1}{z^2 \sqrt{g(z)}} dy
\end{equation}

The Dirac equation that follows from \eqref{diracac} is simply given by
\begin{equation}\label{diraceq1}
 \bigg(\Gamma^{a}e_a^{M} {\mathcal D}_M - m \bigg) \psi = 0.
\end{equation}
Here $e_a^{M}$ are related to the tetrad defined in \eqref{tetrad} through the relations
\begin{equation}
 e^N_b = e^a_M g^{MN} \eta_{ab}
\end{equation}
and are given by
\begin{equation}\label{tetradinv}
 e_{\bar{t}}^t =  \frac{1}{\sqrt{f(z)}}; ~~e_{\bar{z}}^z =  \frac{1}{\sqrt{g(z)}};~~e_{\bar{\theta}}^\theta =  \frac{1}{\sqrt{k(z)}};~~e_{\bar{x}}^x = z = e_{\bar{y}}^y.
\end{equation}

Now let us make the following ansatz for our spinor field
\begin{equation}
 \psi = \frac{z}{\left(f(z)k(z)\right)^\frac{1}{4}} e^{i\left(-\omega t + p \theta + k_1 x + k_2 y\right)} \Psi.
\end{equation}
Here $p$ denotes the momentum of the fermion along the $\theta$ circle. Note that due to the anti-periodic boundary condition of the fermions
$p$ is an half integer times $\pi$. With this redefinition the Dirac equation \eqref{diraceq1} reduces to
\begin{equation}\label{diraceq2}
 \bigg( \frac{1}{\sqrt{g(z)}} \Gamma^{\bar{r}} \partial_r  - m + i z K_{\mu} \Gamma^{\bar{\mu}} \bigg) \Psi =0
\end{equation}
where $\mu$ runs over only boundary coordinates. In \eqref{diraceq2} $K_{\mu}$ is defined to be the following quantity
\begin{equation}
 K_{\mu} = \{-\frac{1}{z \sqrt{f(z)}} \bigg( \omega + q h(z) \bigg),p, k_1, k_2 \}
\end{equation}
Now exploring the rotational symmetry in the $xy$-plane we set $k_2=0$ and $k_1 = k$.

We would also work in the large charge large mass approximation. Also, for capturing a nontrivial dynamics we scale the
frequency $\omega$ and momentum $k$ so that they are same order as the mass and the charge. For convenience we choose a
large number $\gamma$ and keep it explicitly in our equations by considering the following scaling
\begin{equation}
 m \rightarrow \gamma ~m, ~~q \rightarrow \gamma, ~~\omega \rightarrow \gamma ~\omega,~~ k \rightarrow \gamma ~ k.
\end{equation}
Note that for simplicity we shall not scale the $\theta$ circle momentum. This would amount to neglecting the term proportional to $p$ \eqref{diraceq2}.
This assumption physically means that we are ignoring any dependence of our spinor field on the $\theta$ coordinate. This assumption is also
justified from the fact that from the boundary point of view we are interested in the $2+1$ dimensional physics that is obtained after the
Scherk-Schwarz compactification (on the $\theta$-circle) has been performed.

Once the term proportional to $p$ is dropped assuming large $\gamma$, there are only 3 Gamma matrices that are involved in
\eqref{diraceq2}. Following \cite{Hartnoll:2011dm} we make the following choice of the Gamma matrices
\begin{equation}
 \Gamma^{\bar{r}} = \left( \begin{array}{cc} \sigma_3 & ~0 \\ ~0 & \sigma_3 \end{array} \right), ~~
 \Gamma^{\bar{t}} = \left( \begin{array}{cc} i \sigma_1 & ~0 \\ ~0 & i \sigma_1 \end{array} \right), ~~
 \Gamma^{\bar{x}} = \left( \begin{array}{cc} - \sigma_2 & ~0 \\ ~0 &  \sigma_2 \end{array} \right). ~~
\end{equation}
With this choice of the Gamma matrices if we take
$$\Psi = \left( \begin{array}{c} \Psi_1 \\ \Psi_2 \end{array} \right),$$
then from \eqref{diraceq2} it follows that the equations for $\Psi_1$ and $\Psi_2$ decouple and without loss of generality
we can focus on one of the equations given by
\begin{equation}\label{diraceq3}
 \bigg( \frac{1}{\sqrt{g(z)}} \sigma_3 \partial_z  + \gamma \bigg(- m  - i z k \sigma_2  + \frac{1}{\sqrt{f(z)}} (\omega + A_t) \sigma_1 \bigg) \bigg) \Psi_1 = 0.
\end{equation}
Note that since equation \eqref{diraceq3} is a equation for a two component spinor $\Psi_1$ and it is identical to equation (9) in \cite{Hartnoll:2011dm}.
Therefore although our analysis is very similar to that in \cite{Hartnoll:2011dm} but our boundary conditions will be the different which will turn out to be
extremely crucial. Now since $\Psi_1$ is a two component spinor let us consider
$$ \Psi_1 = \left( \begin{array}{c}  \Phi_1 \\ \Phi_2 \end{array} \right) .$$
Then the two equations that follows from \eqref{diraceq3} are
\begin{equation}
 \begin{split}
  \frac{1}{\sqrt{g(z)}} \partial_z \Phi_1 + \gamma \bigg(- m \Phi_1  - z  k \Phi_2 + \frac{1}{\sqrt{f(z)}} (\omega + A_t) \Phi_2 \bigg)  &= 0, \\
  -\frac{1}{\sqrt{g(z)}} \partial_z \Phi_2 + \gamma \bigg(- m \Phi_2 + z k \Phi_1 + \frac{1}{\sqrt{f(z)}} (\omega + A_t) \Phi_1 \bigg)  &= 0.
 \end{split}
\end{equation}
Eliminating $\Phi_2$ from the above equations and keeping only leading order terms in $\gamma$
(which is justified if we are not concerned about the precise energies of the low lying states, see \cite{Hartnoll:2011dm} for details)
we have
\begin{equation}\label{scheq}
 \partial_z^2\Phi_1- \gamma^2 V(z) \Phi_1 =0.
\end{equation}
where the Schr\"odinger potential is given by
\begin{equation}\label{pot}
 V(z) = g(z) \bigg( m^2 + z^2 k^2 - \f{(\omega + h(z))^2}{f(z)}\bigg).
\end{equation}
We plot this potential in fig.\ref{fig:pot} for the solution shown in fig.\ref{fig:ssnum} for $\omega=0$. We now proceed to perform a WKB study of the equations \eqref{scheq} and \eqref{pot}.
It is clear from the plot in fig.\ref{fig:pot} that if $k \gg 0$, then the potential is always positive (see the green lines in fig.\ref{fig:pot} for example).
This regime of momentum and frequency range is uninteresting from the point of view of capturing the gapless excitations about the fermi surface.
From fig.\ref{fig:pot} we also see that for generically (even away from the $\omega =0$ considered for these plots)
the potential has a single turning point which we call $z_{\star}$, where the potential $V(z)$ linearly vanishes.
Note that for $\omega = 0$ and $k=0$ (the blue lines in fig.\ref{fig:pot}), $z_{\star}$ is simply the radius of our soliton star. There are a small number of cases
for which there are two turning points (for instance the yellow line for $m=0.33, \beta =70.43, h_0 =0.7$). However, the final result of
the WKB analysis (existence of fermi surface poles) is same as the case with the single turning point. Therefore, we present here
the details of the WKB analysis for the case with a single turning point.

\FIGURE{
\centering
\includegraphics[width=\textwidth]{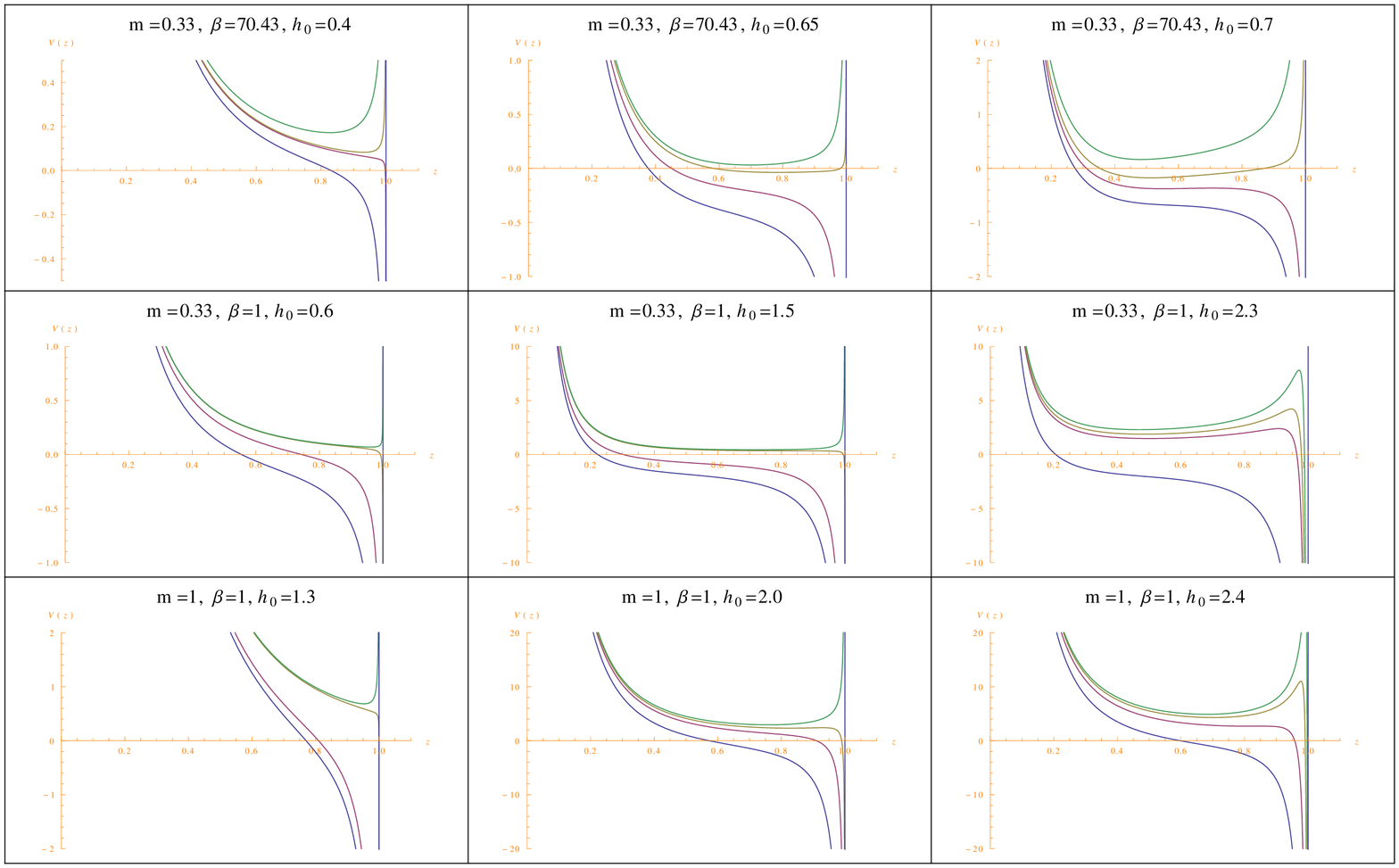}
\caption{Plot of potential $V(z)$ as a function of z for the soliton star in the $\omega = 0$ limit. We plot it for various values of of parameters as indicated in
the plots. The blue line represents the $\omega =0, k=0$ plot in all the cases; therefore the turning point of this blue plot represents the radius of the star. We
increase the momentum $k$ along the red, yellow and green line (in that order).}
\label{fig:pot}
}

For the generic case of a single turning point (at $z_{\star}$) in fig.\ref{fig:pot}, we see that $V(z) < 0$ when $z \in [z_{\star}, 1]$ while
near the boundary of AdS for $z \in [0, z_{\star}]$, we have $V(z) >0$. For convenience let us define the following
quantities
\begin{equation}
 X(z) = \gamma \int_{z}^{z_{\star}} dz \sqrt{V(z)} ;~~ Y(z)= \gamma \int_{z_{\star}}^z dz \sqrt{-V(z)}; ~~ \hat{X} = X(0); ~~ \hat{Y} = Y(1).
\end{equation}
Then in terms of $X$ and $Y$ we can write the following most general solution of \eqref{scheq}
\begin{equation}
 \begin{split}
 \Phi_1(z) &= {\mathcal A} ~e^{i Y(z)} + {\mathcal B} ~e^{-i Y(z)}, ~~~\text{for} ~z \in [z_{\star},1] \\
 \Phi_1(z) &= {\mathcal C} ~e^{X(z)} + {\mathcal D} ~e^{-X(z)}, ~~~\text{for} ~z \in [z_{\star},1] .
 \end{split}
\end{equation}
In order to be consistent with the anti-periodic condition of the fermion about the $\theta$-circle we would
like to the impose vanishing of the fermionic field at the IR cut off $z=1$.Thus demanding
$\Phi_1(z)|_{z=1} = 0$, yields
\begin{equation}\label{nbdc}
 \frac{{\mathcal A} + {\mathcal B}}{{\mathcal A} - {\mathcal B}} = -i \tan{\hat{Y}}
\end{equation}
Further we impose matching conditions at the turning point $z_{star}$. Using the same procedure as explicated in
appendix A of \cite{Hartnoll:2011dm} we find
\begin{equation}\label{mbdc}
\begin{split}
 {\mathcal C} &= ({\mathcal A} + {\mathcal B}) \cos{\frac{\pi}{4}} - i ({\mathcal A} - {\mathcal B}) \sin{\frac{\pi}{4}} \\
 {\mathcal D} &= \frac{1}{2}\bigg( ({\mathcal A} + {\mathcal B}) \sin{\frac{\pi}{4}} + i ({\mathcal A} - {\mathcal B}) \cos{\frac{\pi}{4}} \bigg) \\
\end{split}
\end{equation}
Now we know that the poles of the boundary Greens functions occurs where the function $\Phi_1$ vanishes at the boundary. This
conditions implies that the poles of boundary Greens functions occurs when
\begin{equation}\label{poles}
 \frac{C}{D} = e^{-2 ~\hat{X}}
\end{equation}
 Now using \eqref{nbdc} and \eqref{mbdc} in \eqref{poles} we have the following condition
\begin{equation}
 \tan{\bigg( \hat{Y} + \frac{\pi}{4} \bigg)} + 2 e^{-2 \hat{X}} = 0.
\end{equation}
In the large $\gamma$ limit this condition implies
\begin{equation}\label{poles2}
 \hat{Y} + \frac{\pi}{4} = n \pi - 2 e^{-2\hat{X}}  \Rightarrow  \hat{Y} \approx n \pi.
\end{equation}
Note in that the frequencies solved from this equation do not have any imaginary part (which is unlike case II in \cite{Hartnoll:2011dm}).
This is expected since we do not have any horizon in our soliton star geometry and therefore there is no dissipation.
Also note that in the large $\gamma$ approximation that we are working, the exponential terms are very small and therefore
may be legitimately neglected.

We know that in a fermi liquid there are low energy excitations near the fermi-surface momentum $k_F$, which implies
the existence of a pole in the Green's function at $\omega = 0$ and  $k=k_F$.  Thus in our case we see there are many fermi surfaces
corresponding to fermi momentum $k_F^{(n)}$ described by the equations
$$\hat{Y}(\omega = 0, k= k_F^{(n)}) = n \pi.$$
It is particularly interesting to note that the existence of large number of well defined fermi-surface is
similar to that observed in the gauge theory with adjoint fermions in the free limit discussed in appendix \ref{planarfermion}.

\subsection{Luttinger theorem for the soliton star}

According to the Luttinger theorem \cite{Luttinger:1960zz,2000PhRvL..84.3370O,Huijse:2011hp} the total charge density of the system is proportional to the volume enclosed by the fermi-surface times the charge of the fermionic species under consideration. Our soliton star lacks the presence of any horizon, and all the charge in the system is carried by the fermions themselves. Therefore, Luttinger theorem is expected to hold in our system as in the electron star case \cite{Hartnoll:2011dm}.

On the other hand, in the gravity dual, the following relation has been found \cite{Hartnoll:2011fn,Sachdev:2011ze} (in $d+1$ dimensional bulk space-time)
\begin{equation}
 \frac{q}{(2\pi)^{d-1}} V_F = \rho - A, \label{adsl}
\end{equation}
where $q$ is the charge of the fermion, $\rho$ is the total charge density of the system, and $V_F$ is the total volume enclosed by all the fermi-surfaces.
Here $A$ is the anomalous term which, if non-zero for a system, implies that the Luttinger theorem is invalid in that system. In terms of the bulk fields it
is given by
\begin{equation}
 A = \bigg( - \frac{1}{\lambda^2} \sqrt{-g} F^{rt}(z)|_{z=z_{IR}} \bigg).
\end{equation}
where $\lambda$ is the gauge coupling and $z_{IR}$ is the end point of the IR geometry. For example in the case of a black hole, $z_{IR}$
will be the horizon radius, while for the electron star of \S\ref{sec:elecstar} it is
$z_{IR} = \infty$. This relation (\ref{adsl}) has been proved under very general assumptions (which are valid for our soliton star system) in \cite{Iqbal:2011bf} (see also \cite{Hartnoll:2011pp}
for an analysis where both confined and deconfined fermions coexist).

For our case $z_{IR} =1$ and corresponds to the IR tip of the soliton. From the behavior of the solution star near the IR tip
(discussed in \S\ref{sec:persol}), it follows that the $zt$-component of the gauge field strength $F^{zt}$ (or the $z$-component of the bulk electric field)
vanishes since the time component of the gauge field goes to a constant near the tip of the soliton star. Thus with $A=0$ in this soliton star system, Luttinger theorem
clearly holds.

\section{Conclusions and Discussions}\label{sec:conclude}

In this paper we studied a holographic model of confined fermi liquid by putting fermions
in the AdS soliton geometry. We numerically solved this back-reacted problem and studied its
phase structure when we change the charge density at zero temperature. We also constructed
perturbative solutions analytically and confirmed that they match with the numerical ones.
As we summarized in Fig.\ref{fig:threephase} and Fig.\ref{fig:phased} in section \ref{compa} there are three different cases depending on the values of the parameters $m$ and $\beta$.

The case 1 corresponds to the large $\beta$ and small $m$. This corresponds to a
system with a lot of fermi surfaces and the back-reaction of fermi liquids in the gravity dual can be kept small compared with other two cases. Therefore we have stable soliton star solutions even for large values of the charge density. We made an interesting observation that these solutions approach those of electron stars away from the tip.
The case 2 is realized when both $\beta$ and $m$ are small. In this case, due to the back-reaction of fermi liquids in the bulk gravity, the soliton star gets unstable when
the charge density is very large. It decays into the extremal charged black holes or the
electron stars via a first order phase transition. 
For large values of $m$, the soliton star gets unstable at a relatively small
charge density and the numerical calculations of the energy and charge density predicts that
the soliton star will decay into a new solutions instead of the extremal charge black hole or
the electron stars. This is the case 3. The new solutions are expected to be inhomogeneous. It will be a very interesting future problem to construct such new inhomogeneous solutions and to understand the nature of this phase transition.

We also performed a probe fermion analysis and show that there are many fermi surfaces, which are expected to behave like Landau's fermi liquids. The understanding of all of these properties in the gravity dual from the dual CFT viewpoint is an intriguing future problem. Also it will be another interesting future direction to see how these phase structures change at finite temperature.

\vskip1cm
\noindent
{\bf \large{Acknowledgement}}\\
We are very grateful to S. Hartnoll, R. Loganayagam, S. Minwalla, and S. Sachdev for careful reading of the draft of this paper and valuable comments. 
We would like to thank T. Nishioka, K. Papadodimas and S. Ryu for useful discussions. T.T. thank very much KITP and the organizers of the KITP program 
``Holographic Duality and Condensed Matter Physics'', where
a part of this work has been conducted. N.O. is supported by the postdoctoral fellowship
program of the Japan Society for the Promotion of Science (JSPS) and
partly by JSPS Grant-in-Aid for JSPS Fellows No. 22-4554. The work
of T.T. is also supported in part by JSPS Grant-in-Aid for
Scientific Research No.20740132, and by JSPS Grant-in-Aid for
Creative Scientific Research No.\,19GS0219. We are supported by
World Premier International Research Center Initiative (WPI
Initiative), MEXT, Japan.


\appendix

\section{$O(N^2)$ Fermi Surfaces from $SU(N)$ Yang-Mills}\label{planarfermion}

Let us consider an adjoint fermion coupled to a $SU(N)$ gauge
field in 4 dimensions with a global $U(1)$ chiral symmetry. In order to
find an evidence that this system have $O(N^2)$ Fermi surfaces,
we wish to evaluate the partition function of this system and hence compute
the specific heat in the free limit. Possibilities that gauge non-singlet sectors of
fermions make fermi surfaces in addition to the gauge singlet one have been already discussed in \cite{Huijse:2011hp} in different setups.

We shall closely follow the basic set up in
\cite{Aharony:2003sx,Yamada:2006rx}. The grand canonical partition
function is given by
\begin{equation}
 Z(\beta, \mu) = Tr \bigg( \exp (- \beta (H-\mu Q)) \bigg),
\end{equation}
$Q$ being the chiral charge operator which is a conserved $U(1)$
global charge for this system. Here $\beta = 1/T$ and $\mu$ are
respectively the inverse temperature and the chemical potential for
the $U(1)$ global symmetry. Let us consider this theory on $S^3
\times S^1$ with $R$ being the $S^3$ radius and $\beta$ being the
$S^1$ radius. Ultimately we shall be interested in the large $R$
limit in which the sphere tends to flat $R^3$. Here we shall set
$R=1$, so this flat space limit will correspond to $\beta \ll 1$,
that is when the radius of the time circle is very small compared to
the $S^3$ radius.

The partition function may also be written in terms of a functional
path integral
\begin{equation}
 Z = \int {\mathcal D} A_{\mu} {\mathcal D} \psi e^{-S[A_{\mu}, \psi]}.
\end{equation}
We have to impose periodic boundary condition for the gauge field
and anti-periodic boundary conditions for the fermion along the time
circle. In the free limit of the theory all the heavy modes of the
fields may be integrated out (by evaluating one loop vacuum
diagrams). However, the zero mode of the time component of the gauge
field $A_0$ which is strongly coupled even in the free limit has to
be treated exactly. Thus, the process of integrating out the heavy
modes yields a quantum effective action for this zero mode which may
be written in terms a matrix integral
\cite{Aharony:2003sx,Yamada:2006rx} which has the following form
\begin{equation}\label{zseff}
 Z = \int dU e^{-S_{eff}(U)},
\end{equation}
with $U$ being a $SU(N)$ matrix. The effective action $S_{eff}(U)$
may also be evaluated in the free limit by explicitly counting the
gauge invariant states in the free theory on $S^3 \times S^1$ and
then projecting onto the singlet sector so as to ensure Gauss law
constraint. This effective action is evaluated to be
\cite{Aharony:2003sx,Yamada:2006rx}
\begin{equation}\label{seff}
 S_{eff}(U) = -\sum_{n=1}^{\infty} \frac{1}{n} \left( z_v(x^n) + (-1)^{n+1} z_f(x^n,\mu) \right) \bigg(Tr(U^n)Tr(U^{\dagger ~n}) -1\bigg),
\end{equation}
where $x = e^{-\beta}$ and  $z_v$, $z_f$ are respectively the single
particle partition functions for the gauge field and the fermionic
field, which are given by
\begin{equation}
 \begin{split}
  z_v(x) &= \frac{6x^2 -2x^3}{(1-x)^3}, \\
  z_f(x,\mu) &= \frac{2 x^{\frac{3}{2}}}{(1-x)^3} \bigg( x^{\mu} + x^{-\mu}\bigg)
 \end{split}
\end{equation}
Note that for concreteness we have considered a chiral fermion in writing down this single particle 
partition function $z_f(x,\mu)$. However the final result of this appendix also holds for a
Dirac fermion.

Now we intend to perform the integral in \eqref{zseff} by rewriting
it in terms of the eigenvalues $(e^{i \theta})$ of the matrix $U$.
The integration measure $dU$ is given by the Haar measure of
$SU(N)$. Introducing a suitably normalized eigenvalue distribution
function $\rho(\theta)$, the effective action in \eqref{seff} may be
written as \cite{Aharony:2003sx,Yamada:2006rx}
\begin{equation}\label{seffrho}
 S_{eff} (\rho) = N^2 \sum_{n=1}^{\infty} V_n |\rho_n|^2,
\end{equation}
where $\rho_n = \int_{-\pi}^{\pi} d\theta \rho(\theta)~ \cos{n
\theta}$, are the Fourier coefficients of the eigenvalue
distribution. Here $V_n$ is given by
\begin{equation}
 V_n = \frac{1}{n} \bigg(1-z_v(x^n)-(-1)^{n+1} z_f(x^n, \mu) \bigg).
\end{equation}

Note that until now the effective matrix integral that we have found
is exact in $N$. But now we wish to take a large $N$ limit, and
evaluate the integral \eqref{zseff} in the saddle point
approximation. In this limit there are essentially two saddle
points. In \eqref{seffrho}, when the $V_n$ are all positive they act
as a repulsive potential for the eigenvalues, while when they are
negative they act as attractive potential. As a result, for positive
$V_n$, a saddle point is obtained where the eigenvalues are
uniformly distributed, that is $\rho_{n \geq 1} = 0$. This
corresponds to the confined phase. On the other hand, when $V_n$ is
no longer positive the action is minimized by a tightly clustered
configuration of eigenvalues and in the large $N$ limit this is
given by the saddle point $\rho_n = 1$, for all $n$. This
corresponds to the deconfined phase.

As we have emphasized for the confined phase to exist we must have
$V_n > 0$ for all $n$. This implies that $V_n =0$ will provide us
with the line of phase transition in the $\mu T-$plane. However,
since $z_v$ and $z_f$ are monotonically increasing functions with $0
\leq x < 1$, the most stringent condition is obtained when $n=1$.
Therefore the equation of the curve separating the confined and
deconfined phases in the $\mu T-$plane is given by the equation
\begin{equation}
 z_v(x)+ z_f(x, \mu) = 1.
\end{equation}
This curve has been plotted in fig.\ref{fig:muTplane}.

\FIGURE{
\centering
\includegraphics[width=0.7\textwidth]{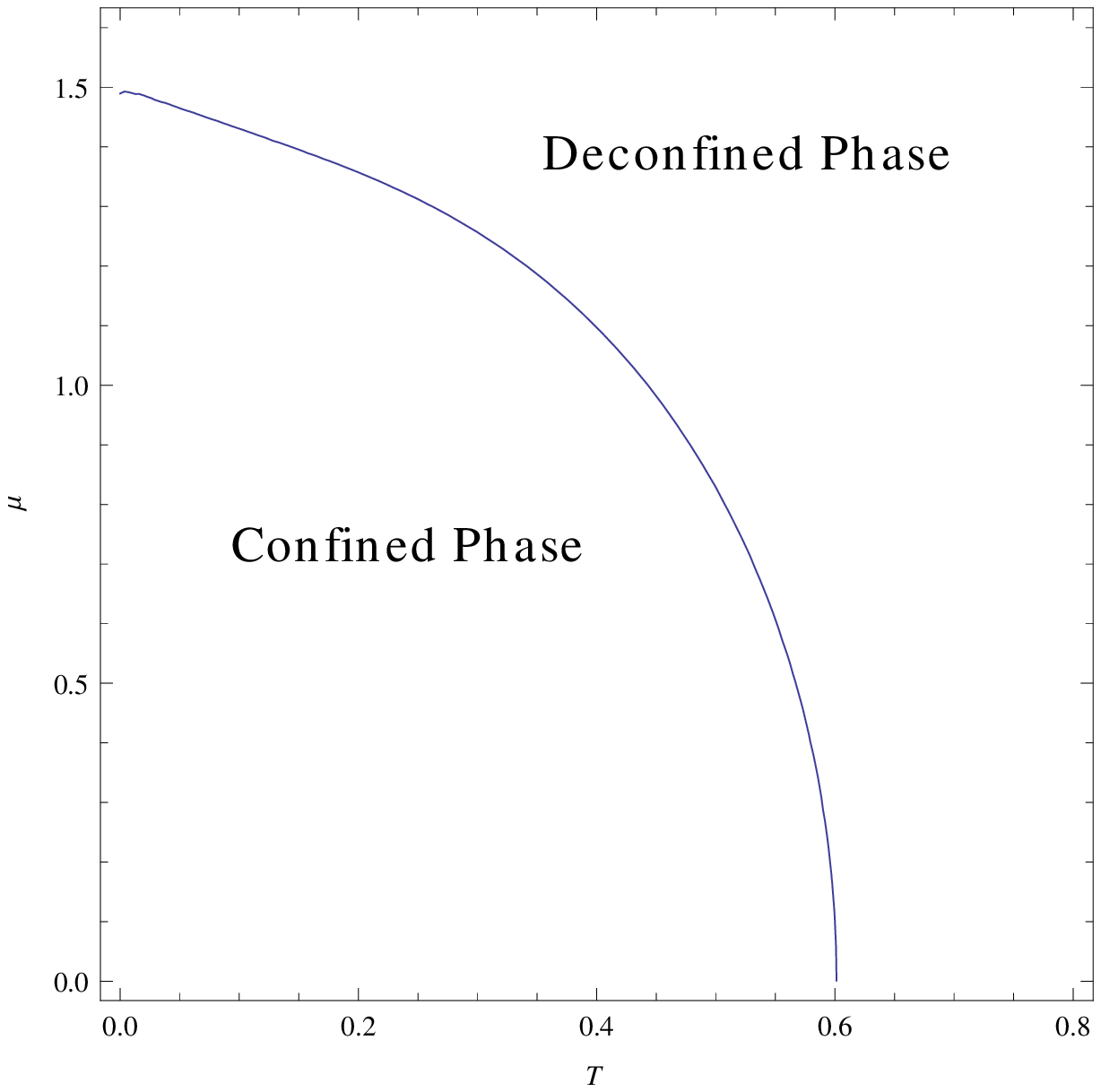}
\caption{Confinement deconfinement phase transition curve in the
$\mu T$-plane} \label{fig:muTplane}
}

In this discussion we would particularly like to focus on the
deconfined phase in the limit $\mu \gg 1$, $\beta \ll 1$ so that
$\beta \mu \gg 1$. In this limit, since the radius of the thermal
circle becomes small compared to that of the $S^3$ radius (which has
been set to 1), we we can approximate $S^3$ by flat space. The other
condition $\beta \mu \gg 1$ ensures that we are in the deconfined
phase. Since we are in the deconfined phase we have $\rho_n = 1$.
The effective action \eqref{seffrho} in this phase is then given by,
\begin{equation}\label{seffdc}
 S_{eff}\bigg|_{\text{deconfined}} = N^2 \sum_{n=1}^{\infty} V_n (\mu \beta \gg 1) = N^2 \sum_{n=1}^{\infty} \frac{1}{n} \bigg(1-z_v(x^n)-(-1)^{n+1} z_f(x^n, \mu) \bigg).
\end{equation}
In the sum in \eqref{seffdc} there are two contributions. One
contribution comes from the gauge bosons through $z_v$ and the other
contribution comes from the fermions through $z_f$. The bosonic
contribution is independent of $\mu$ which a consequence of the fact
that the gauge bosons are uncharged under the global $U(1)$
symmetry. As a consequence of the this fact, in the limit ($\beta
\mu \gg 1$) under consideration, the leading contribution comes from
the fermions. The first term in \eqref{seffdc} merely contributes to
a constant additive shift to the effective action and may be
absorbed in to a normalization of the partition function.

In the large N limit, the grand potential in this deconfined phase
saddle point is given by
\begin{equation}
 \Omega = - \frac{1}{\beta} \ln Z = \frac{1}{\beta} S_{eff}\bigg|_{\text{deconfined}}
\end{equation}
Then using \eqref{seffdc} and the approximations explained above the
grand potential reduces to,
\begin{equation}
\begin{split}\label{omega}
 \Omega &= - \frac{N^2}{\beta} \sum_{n=1}^{\infty} \frac{1}{n} \bigg((-1)^{n+1}  \frac{2 x^{\frac{3n}{2}}}{(1-x^n)^3} \bigg( x^{n \mu} + x^{-n \mu}\bigg)\bigg) \\
        &= - \frac{2 N^2}{\beta} \sum_{m=0}^{\infty} \left( \begin{array}{c} m+2 \\ m\end{array} \right) \bigg( \ln (1-x^{m+\frac{3}{2}+ \mu})+\ln (1-x^{m+\frac{3}{2}-\mu}) \bigg)
\end{split}
\end{equation}
In the limit that we are working in the sum in \eqref{omega}
receives primary contribution from large values of $m$, when the sum
may be replaced by an integral. This approximation essentially
corresponds to the fact that in the flat space approximation that we
are working in the discrete spectrum on the sphere becomes
continuous. Thus we have
\begin{equation}
  \Omega = -\frac{2 N^2}{\beta}\int_0^{\infty} dm ~m^2 \bigg( \ln (1-x^{m+\frac{3}{2}+ \mu})+\ln (1-x^{m+\frac{3}{2}-\mu}) \bigg),
\end{equation}
Again neglecting the $3/2$ in comparison to the large value of $\mu$
and performing an integration by parts we obtain
\begin{equation}
  \Omega = - \frac{2 N^2}{3} \int_0^{\infty} dm  \bigg( \frac{m^3}{e^{\beta(m-\mu)}+1}+\frac{m^3}{e^{\beta(m+\mu)}+1}) \bigg),
\end{equation}
which in the limit $\beta \mu \gg 1$ may be approximated by
\footnote{Here we have used the fact that in the $\mu/T \gg 1$ limit
we have
\begin{equation}
 \int_0^{\infty} \frac{f(m)}{e^{\frac{m-\mu}{T}}+1} \approx \int_0^{\infty} f(m) dm + \frac{\pi^2}{6} T^2 f'(\mu) + {\mathcal O} (T^4)
\end{equation}
}
\begin{equation}
  \Omega(\mu,T) \approx \Omega_0(\mu) - \frac{2 N^2 \pi^2}{3} T^2 \mu^2 ,
\end{equation}
Now to calculate the specific heat we use the following
thermodynamic relations
\begin{equation}
 S = - \big(\partial_T \Omega \big)_{V,\mu},~N = - \big(\partial_\mu \Omega \big)_{V,T},~C_v = \big(\partial_T E \big)_{V,N} = T \big(\partial_T S \big)_{V,N}
\end{equation}
Note that we hold the volume of the space ($S^3$) fixed everywhere
and therefore it does not appear explicitly anywhere and we need not
worry about it separately. Also it follows from the above
relations that holding N fixed is the same as holding $\mu$ fixed
upto the leading order; therefore to leading order in $\mu/T$ we have
\begin{equation}
 C_v = \frac{4 N^2 \pi^2}{3} T \mu^2
\end{equation}
Note the linear dependence of specific heat on the temperature $T$
which is a standard well known result. Also note that the specific
heat is proportional to $N^2$. Since each fermi-surface has the chemical potential
of order $\mu$ and contributes ${\mathcal O}(T)$ to the specific heat,
this implies that there are $N^2$ number of fermi-surfaces for the theory we consider here.
A straightforward generalization of this argument to arbitrary dimensions shows that the 
main result of this appendix (presence of $N^2$ number of fermi surfaces) continues to hold 
in the generalized case.

\section{Scaling Argument}\label{resc}

We start with the Maxwell action \eqref{EMth} coupled to the energy momentum tensor \eqref{EMT} of Fermi liquids with the profile \eqref{rho} and \eqref{sig}.
We define the metric with the AdS radius $L$ and the gauge potential by the
ansatz
\ba
&& ds^2=L^2\left(-f(z)dt^2+g(z)dz^2+k(z)d\theta^2+z^{-2}(dx^2+dy^2)\right),\no
&& A=\f{eL}{\kappa}h(z)dt.
\ea

We defined the rescaled quantities $\hat{\beta},\hat{m},\hat{p}$ and $\hat{\rho}$ as follows
\be
p(z)=\f{\hat{p}(z)}{L^2\kappa^2},\ \ \ \rho(z)=\f{\hat{\rho}(z)}{L^2\kappa^2}, \ \ \
\beta=\f{\kappa^3}{e^5L^2}\hat{\beta},\ \ \ m=\f{e}{\kappa}\hat{m}. \label{estate}
\ee
If we substitute \eqref{estate} into the equations of motion, the dependence on $e,\kappa$ and
$L$ drops out. Therefore in the bulk of this paper, we can set $e=\kappa=L=1$ without losing
generality. If we would like to
know the results for general values of $e,\kappa$ and $L$, then we just need to replace
$(\beta,m,p,\rho)$ in the context of this paper
with $(\hat{\beta},\hat{m},\hat{p},\hat{\rho})$. It is clear from this analysis that the
normalized parameters $\hat{m}$ and $\hat{\beta}$ are not only proportional to the mass and
species of bulk fermion but also depend on the charge $e$ of the bulk fermion.

\section{The functions in the perturbative soliton star solution} \label{perfunc}

The functions constituting the linear order corrections in $\beta$ to the  solution inside the star
{\small
\begin{equation}
 \begin{split}
  f^{(1)}_{in}(z) &= \frac{h_0 }{48 z^2}\bigg(2 \left(3 h_0^2
   \left(h_0^2+8\right)-16\right) \log \left(1-z^4\right)\\&-3 h_0^2
   \left(h_0^2 (8 z+\pi -8+\log (64))-8 \pi +8 \log (64)\right)+4
   \left(3 h_0^2 \left(h_0^2-8\right)-16\right) \tan ^{-1}(z) \\& +4
   \left(3 h_0^2 \left(h_0^2+8\right)-16\right) \tanh ^{-1}(z)+16
   (\pi +\log (64))\bigg)\\
  g^{(1)}_{in}(z) &= \frac{1}{360
   z^2 \left(z^4-1\right)^2} \bigg( -10 h_0 \left(3 h_0^2 \left(h_0^2+8\right)-16\right)
   z^4 \left(\log \left(z^2+1\right)+2 \log (z+1)\right) \\& +z^3 \left(3
   h_0^5 z (6 (4 z-4+\log (32))+5 \pi )-40 h_0^3 (z (-8+3 \pi -18
   \log (2))+8) \right. \\& \left.-80 h_0 z (\pi +\log (64))-256 z\right)+20 h_0
   \left(-3 h_0^4+24 h_0^2+16\right) z^4 \tan ^{-1}(z)+256 \bigg)\\
  k^{(1)}_{in}(z) &= \frac{1}{2880 z^2}\bigg(-3 h_0^5 \left(48 z^5+(5 \pi -52) z^4+40 z+5 \pi -36\right) \\& +40
   h_0^3 \left((4+3 \pi ) z^4-8 z^3+3 \pi +4\right) \\& +10 h_0
   \left(z^4+1\right) \left(3 \left(h_0^2+8\right) h_0^2
   \left(\log \left(\frac{1}{8} \left(z^2+1\right)\right)+2 \log
   (z+1)\right) \right. \\& \left. +\left(6 h_0^2 \left(h_0^2-8\right)-32\right) \tan
   ^{-1}(z)+8 \left(-2 \log \left(z^2+1\right)-4 \log (z+1)+\log
   (64)\right)\right) \\ & +80 h_0 \left((4+\pi ) z^4+8 z+\pi -12\right)-512
   \left(z^4-1\right) \bigg)\\
  h^{(1)}_{in}(z) &= \frac{1}{96 z} \bigg( \left(3 h_0^2 \left(h_0^2+8\right)-16\right) z \left(-4
   \tanh ^{-1}\left(z^2\right)+4 \tanh ^{-1}(z)-\log (4)\right) \\&+\left(\pi
   \left(3 h_0^2 \left(h_0^2-8\right)-16\right)-128\right) z+4
   \left(-3 h_0^4+24 h_0^2+16\right) z \tan ^{-1}(z)+128 \bigg)\\
 \end{split}
\end{equation}
}

The ${\mathcal O} (\beta)$ correction to the star radius is given by,
{\small
\begin{equation}
 \begin{split}
  z_r^{(1)} &= -\frac{m}{96 h_0^2} \left(-8 (h_0-m) \left(3 h_0^5-m^3\right)+\pi
   \left(h_0^2+1\right) \left(3 h_0^4-6 h_0^2
   m^2-m^4\right) \right. \\ &\left. -2 \left(3 h_0^4+6 h_0^2 m^2-m^4\right)
   \left(\log \left(\frac{2
   \left(h_0^2+m^2\right)}{(h_0+m)^2}\right)+h_0^2 \log
   \left(\frac{(h_0+m)^2 \left(h_0^2+m^2\right)}{8
   h_0^4}\right)\right) \right. \\ &\left. -4 \left(h_0^2+1\right) \left(3
   h_0^4-6 h_0^2 m^2-m^4\right) \tan
   ^{-1}\left(\frac{m}{h_0}\right)\right)\\
 \end{split}
\end{equation}
}

The integration constants in the solution outside \eqref{outpsol} is given by
{\small
\begin{equation}
 \begin{split}
  F_1 &= \frac{1}{48} h_0 \bigg( 2 (m-h_0)^3 (3 h_0+m) \log
   \left(\frac{m^4}{h_0^4}-1\right)-3 h_0^4 (-8+\pi +\log
   (64)) \\& -24 h_0^3 m+6 h_0^2 m^2 (\pi -6 \log (2))+2 m^2
   \left(m^2-6 h_0^2\right) \left(-\log \left(h_0^2+m^2\right)-2
   \log (h_0+m) \right. \\&  \left. +4 \log (h_0)\right)+6 h_0^4 \log
   \left(\frac{(h_0+m)^2
   \left(h_0^2+m^2\right)}{h_0^4}\right)+4 \left(3 h_0^4-6
   h_0^2 m^2-m^4\right) \tan ^{-1}\left(\frac{m}{h_0}\right) \\& +m^4
   (\pi +\log (64)) \bigg)\\
  F_2 &= -\frac{1}{6} h_0 (h_0-m)^3 (3 h_0+m)\\
  K_1 &=\frac{1}{1440} \bigg( -3 h_0^5 (-39+5 \pi +30 \log (2))-120 h_0^4 m+10
   h_0^3 m^2 (1+3 \pi -18 \log (2))\\ &+20 \left(3 h_0^5-6
   h_0^3 m^2-h_0 m^4\right) \tan
   ^{-1}\left(\frac{m}{h_0}\right)+10 h_0 \left(\left(3
   h_0^4+6 h_0^2 m^2-m^4\right) \left(\log
   \left(\frac{m^2}{h_0^2}+1\right) \right. \right.\\&  \left. \left. +2 \log
   \left(\frac{h_0+m}{h_0}\right)\right)-(h_0-m)^3 (3
   h_0+m) \log \left(\frac{m^4}{h_0^4}-1\right)\right) \\&  +5
   h_0 m^4 (-3+\pi +\log (64))+8 m^5 \bigg) \\
  K_2 &=-\frac{i (h_0-m)^3 \left(69 h_0^2+47 h_0 m+24
   m^2\right)}{1440} \\
  H_1 &=-\frac{1}{12} (h_0-m)^3 (3 h_0+m) \\
  H_2 &= \frac{1}{96} \bigg(3 \pi  h_0^4-6 \pi  h_0^2 m^2-2
   (h_0-m)^3 (3 h_0+m) \log \left(\frac{2
   h_0^2}{h_0^2+m^2}-1\right) \\&+4 \left(-3 h_0^4+6
   h_0^2 m^2+m^4\right) \tan
   ^{-1}\left(\frac{m}{h_0}\right)+\left(3 h_0^4+6 h_0^2
   m^2-m^4\right) \left(-4 \tanh ^{-1}\left(\frac{m^2}{h_0^2}\right) \right. \\&  \left. +4
   \tanh ^{-1}\left(\frac{m}{h_0}\right)-\log (4)\right)+8 h_0
   m^3-(8+\pi ) m^4\bigg)\\
 \end{split}
\end{equation}
}

\section{Constants in the analytical solution near the tip of the soliton star}\label{tipfunc}

In this appendix we list the constants in \eqref{tipsol} as determined from the equations of motion

The constant appearing in $f(z)$ function
\begin{equation}
 \begin{split}
  f_1 &= \frac{8 \left(-9 \beta  h_0^5+20 \beta  h_0^3 m^2-15 \beta
   h_0 m^4+4 \left(\beta  m^5-45\right)\right)}{9 \beta  h_0^5-10
   \beta  h_0^3 m^2-15 \beta  h_0 m^4+16 \left(\beta
   m^5-45\right)},\\
 \end{split}
\end{equation}
The constants appearing in $g(z)$ are
\begin{equation} \label{tipsolconst1}
 \begin{split}
  g_{(-1)} &= -\frac{180}{9 \beta  h_0^5-10 \beta  h_0^3 m^2-15 \beta  h_0
   m^4+16 \left(\beta  m^5-45\right)} ,\\
  g_0 &= -\frac{180}{\left(9 \beta  h_0^5-10 \beta
   h_0^3 m^2-15 \beta  h_0 m^4+16 \left(\beta
   m^5-45\right)\right)^3}  \bigg(-1134 \beta ^2 h_0^{10} \\ &+135 \beta ^2 h_0^8
   \left(26 m^2+15\right)-5 \beta ^2 h_0^6 m^2 \left(866
   m^2+1215\right)+1188 \beta  h_0^5 \left(\beta  m^5-45\right) \\ &+75
   \beta ^2 h_0^4 m^4 \left(14 m^2+81\right)-760 \beta  h_0^3 m^2
   \left(\beta  m^5-45\right)+225 \beta ^2 h_0^2 m^6 \left(8
   m^2-9\right)\\ &-2220 \beta  h_0 m^4 \left(\beta  m^5-45\right)+896
   \left(\beta  m^5-45\right)^2\bigg)
 \end{split}
\end{equation}
The constant in $k(z)$ is
\begin{equation}
 \begin{split}
  k_2 &= \frac{720}{\left(\beta  \left(9
   h_0^5-10 h_0^3 m^2-15 h_0 m^4+16
   m^5\right)-720\right)^3} \bigg(\beta  \left(162 \beta  h_0^{10} -225 \beta  h_0^8
   \left(2 m^2+3\right) \right. \\ & \left.+5 \beta  h_0^6 m^2 \left(182 m^2+405\right)-684
   h_0^5 \left(\beta  m^5-45\right)-75 \beta  h_0^4 m^4 \left(10
   m^2+27\right) \right. \\ & \left.+1000 h_0^3 m^2 \left(\beta  m^5-45\right)+675 \beta
   h_0^2 m^6-60 h_0 m^4 \left(\beta  m^5-45\right)-128 m^5
   \left(\beta  m^5-90\right)\right) \\ &-259200\bigg)
 \end{split}
\end{equation}
And the constant in $h(z)$ is
\begin{equation}
 \begin{split}
  h_1 &= -\frac{45 \beta  \left(h_0^2-m^2\right)^2}{9 \beta  h_0^5-10 \beta
    h_0^3 m^2-15 \beta  h_0 m^4+16 \left(\beta  m^5-45\right)} \\
 \end{split}
\end{equation}

\bibliographystyle{JHEP}
\bibliography{Soliton_star}

\end{document}